\documentclass[onecolumn]{autart}
\usepackage{tabularx}

\usepackage{soul}
\setstcolor{red}
\usepackage{float}
 \usepackage[utf8x]{inputenc}
\usepackage{amsmath}
\usepackage{hyperref}
\usepackage{graphicx}
\usepackage{amssymb}
\usepackage{verbatim}
\usepackage{epsfig}
\usepackage[labelfont=bf]{caption}
\usepackage{enumerate}
\usepackage{subfig}
\usepackage{epstopdf}
\usepackage[normalem]{ulem}
\usepackage{booktabs} 
\usepackage{algorithmicx,algorithm}
\usepackage{bbm}
\usepackage{mathtools}
\usepackage{tabularx}
\mathtoolsset{showonlyrefs}
 \usepackage{xcolor}
 \newcommand\blfootnote[1]{%
  \begingroup
  \renewcommand\thefootnote{}\footnote{#1}%
  \addtocounter{footnote}{-1}%
  \endgroup
}

\begin{document}
\begin{frontmatter}
\title{PID passivity-based droop control of power converters:\\ Large-signal stability, robustness and performance}
 
\thanks[footnoteinfo]{Corresponding author: Daniele Zonetti. email: daniele.zonetti@gmail.com}

\author[DZ]{Daniele Zonetti\thanksref{footnoteinfo}},    
\author[GB]{Gilbert Bergna-Diaz},       
\author[RO]{Romeo Ortega},
\author[NM]{Nima Monshizadeh}

\address[DZ]{CITCEA-UPC, Polytechnical University of Catalonia, Barcelona, Spain}
\address[GB]{Department of Electric Power Engineering, Norwegian University of Science and Technology, 7491 Trondheim, Norway}             
\address[RO]{Departamento Acad\'{e}mico de Sistemas Digitales, ITAM, 01080 Ciudad de M\'exico, Mexico}        
\address[NM]{Engineering and Technology Institute, University of Groningen, 9747 AG, Groningen, The Netherlands}        


\renewcommand{\thefootnote}{\textit{\alph{footnote}}}


\begin{abstract}We present a full review of PID passivity-based controllers (PBC) applied to power electronic converters, discussing limitations, unprecedented merits and potential improvements in terms of large-signal stability, robustness and performance. We provide four main contributions.  The nominal case is first considered and it is shown---under the assumption of perfect knowledge of the system parameters---that the PID-PBC is able to guarantee global \textit{exponential} stability of a desired operating point for any positive gains.  Second, we analyze robustness of the controller to parameters uncertainty for a specific class of power converters, by establishing precise stability margins. Third, we propose a modification of the controller by introducing a leakage, in order to overcome some of the intrinsic performance and robustness limitations.  Interestingly, such controller can be interpreted at steady-state as a droop between the input and the passive output, similar to traditional primary controllers. Fourth, we robustify the design against saturation of the control input via an appropriate monotone transformation of the controller. The obtained results are thoroughly discussed and validated by simulations on two relevant power applications: a dc/dc boost converter and an HVDC grid-connected voltage source converter.
\end{abstract}
\end{frontmatter}

\blfootnote{\textbf{Abbreviations:} mPLID, monotone proportional leaky integral derivative; PBC, passivity-based control; GAS, global asymptotic stability; GES, global exponential stability; ZI, constant impedance, constant current; 2L-VSC, two-level voltage source converter; HVDC, high-voltage direct-current.}

\section{Introduction}
\subsection{Motivation}
In the past decade the more recent advances in power electronics technologies revolutionized  the way electrical energy is transported and used,  entailing dramatic changes in both the power and automation sectors. The majority of electrical applications runs nowadays on power electronics-based architectures: drive efficiently operating motors are available from 10 W to hundreds of MW. HVDC lines empower transmission of electrical energy over a long distance up to 6 GW, and at a voltage level of almost 1000 kV. Trains, elevators and cranes actuation strongly relies on power electronics. The integration of renewable energy sources, such as wind turbines and photovoltaic panels, to the electric grid is enabled by converters~\cite{bose}.\\
Yet, power converters are highly controllable energy transformation devices for which an accurate control design is required. Proportional Integral Derivative (PID) control is by far the most diffused and universally accepted strategy, whose undisputed success is mostly due to the ease of implementation and to the fact that the design is grounded on linear systems theory, for which powerful analytical tools are readily available. However, because of their switching characteristics, the dynamics of a power converter are essentially nonlinear. As a result, a time-consuming and expensive procedure to tune the gains of the PIDs is required to complete the design and, in view of the wide range of the operating regimes, frequently yield below-par performances~\cite{arun,raviraj}.

PID passivity-based control (PBC) has been proposed as an alternative design based on nonlinear systems theory and is nowadays a widely accepted control strategy for power electronic converters, which has been proven effective in many practical situations~\cite{borja2021pid}. The close relationship between the PID-PBC and the popular Akagi's PQ controller~\cite{akagi2017} has further contributed to its popularity in the power electronics community. The PID-PBC enjoys indeed several features that makes it a serious competitor to traditional PID controllers. First, passivity is an input-output property that is preserved upon interconnection. Hence, stability certificates established by local, passivity-based control designs, immediately extend, under mild assumptions, to the interconnected case. This is particularly relevant for electrical grids, which are---by nature---highly interconnected systems. Second, the approach directly relies on Lyapunov's stability theory. Therefore, the design allows to shape a suitable energy function for the closed-loop system that provides solid ground for the design of higher-level stability-preserving controllers.\\
Nonetheless, although the PID-PBC has been successfully implemented for a broad class of power applications, ranging from smart grids to commercial electronic devices, and in a great variety of operating conditions, many practical control requirements such as minimum performance, robustness to parameters uncertainty, sensing and actuation limitations have been only partially investigated from a theoretical point of view. As a result, practitioners typically implement such a controller either by introducing complicated, \textit{ad hoc} modifications--grounded on power electronics expertise--or by complementing the design via time-consuming tuning procedures, similar to the traditional implementation of PID controllers. Since most of these practical modifications invalidate the theoretical results obtained using passivity arguments, it may be then questioned what is the real benefit of employing a PID-PBC instead of the conventional, ubiquitous, PID controllers. In this paper we provide an answer to this question by means of a full review of PID passivity-based control of power electronic converters. We proceed by rigorously establishing limitations and unprecedented merits of this controller, further proposing appropriate modifications that guarantee improvements in terms of stability, robustness and performance.
 
\subsection{Existing literature} 
The building block for the design of the PID-PBC is the use of appropriate energy-based representations of the power converter models, which lead to a simpler formalization of essential physical concepts such as energy storage and flows, dissipation and interconnection with the external environment. While in this paper we find more convenient to explicitly focus on port-Hamiltonian representations of the power converters~\cite{GEOPLEX,escobar1999hamiltonian}, other energy-based descriptions can be used, such as Euler-Lagrange or Brayton-Moser representations~\cite{ortega2013passivity,jeltsema2003dual}.\\
As already discussed, some limitations have been observed in practical implementation of PID-PBCs and theoretical questions have been raised, concerning four fundamental aspects: performance of the controller; robustness to parametric uncertainty; limited sensing and robustness to saturation of the control input. We review here below contributions available in literature related to these aspects and to PID-PBCs.\\
Rooted on the passivity concepts developed for dc-to-dc converters and based on Euler-Lagrange representations~\cite{sira1997passivity}, the first PI-PBC controller relying on a port-Hamiltonian description was originally presented in~\cite{perez2004passivity} for a broader class of power applications. A similar problem was then recast for the PI-PBC of general nonlinear systems, with application to nonlinear RLC circuits~\cite{jayawardhana2007passivity}. In both cases the PI-PBC was suggested as a, still \textit{linear}, alternative to traditional PI current or voltage controllers, which suffer from well-known internal stability problems~\cite{sira1997passivity,lee2003input}. The use of a derivative action has been traditionally avoided, as this requires the implementation of an additional output filter to attenuate high-frequency noise. However, the use of such a design may considerably improve the system's performances~\cite{li2006pid}. More information on the application of traditional PID controllers to power converters can be found in standard textbooks in power electronics~\cite{erickson2007fundamentals}. Despite its success, the proposed passivity-based design requires the exact knowledge of the system's parameters in order to \textit{a priori} compute the equilibrium to be stabilized---a fact rarely verified in practice. In~\cite{hernandez2009adaptive} the problem of regulation under parametric uncertainty was addressed, and different adaptive schemes were proposed. However, adaptive schemes typically lead to more complicated designs and suffer from other robustness problems, as the non-scalability to large-scale grid applications. In recent works~\cite{Zonetti2015,bergna2018pi}, it was further observed that for some of these applications, such as HVDC transmission system, the PI-PBC may exhibit poor performances. To cope with robustness and performance limitations, it is common practice to design decentralized outer-loop controllers that modify the PI-PBC references---typically provided by an higher-level controller--- to guarantee the overall system's stability. Outer-loop or alternative PI-PBC schemes have been already proposed for general dc/dc converters~\cite{kosaraju2020differentiation} and specific applications: a passivity-based loop was added to a PI-PBC strategy for maximum point power extraction of wind energy conversion systems~\cite{cisneros2016pi}; an alternative PI-PBC, based on the Brayton-Moser representation of a boost converter-based dc microgrids, was designed to guarantee robust output voltage regulation~\cite{cucuzzella}; a droop alike scheme was proposed for voltage source converter-based HVDC transmission systems to speed up convergence~\cite{Zonetti2015}. More recently, we suggested to introduce a leakage in the integral channel to address both performance and robustness issues~\cite{zonetti2020}---an approach that has been adopted also in the context of frequency control in power systems~\cite{mallada} and is pursued in this paper as well.  The problem of limited sensing has been also recently addressed~\cite{jaafar2012pi,bobtsov2020globally}, resulting in finite-time convergent observers to be designed to complement the PI-PBC. The problem of limited actuation, with some notable exception~\cite{guzman2013saturated}, has not been properly investigated for the PI-PBC. Yet, there exist several works that addressed the problem of design of outer-loops guaranteeing that the control input is maintained within prescribed bounds~\cite{rodriguez2000robustly,konstantopoulos2016}.

\subsection{Contributions}
We consider general power converters that can be described by port-Hamiltonian systems without switching sources and assume that the loop is closed by a PID-PBC designed along the lines of~\cite{hernandez2009adaptive} further including a derivative action. Although the choice of not including converters with switching sources---such as the buck and buck-boost converters---may seem arbitrary, it relies on two fundamental observations. First, from a strictly mathematical point of view the switching of an ideal current or voltage source corresponds to a control input that enters linearly in the system dynamics, a fact that considerably simplifies the design.  Second, switching sources are ideal elements that can be always replaced by linear components with arbitrarily fast switching dynamics, which are rigorously analyzed in this paper. Henceforth, it can be shown that all contributions extend to the case of power converters with switching sources with minor differences.  The contributions of our paper are fourfold. 
\begin{itemize}
\item[C1)] We first assume to have full knowledge of the system parameters and, as a result, that the set of assignable equilibria is perfectly known. Hence, upon selection of a desired operating point to be stabilized from such a set, we prove that the PID-PBC ensures that the latter is a globally \textit{exponentially} stable (GES) equilibrium of the closed-loop system. This allows to establish a {worst-case} convergence rate for the PID-PBC, which can be particularly slow for specific power applications. This result should be contrasted with the one reported in~\cite{hernandez2009adaptive} where the strictly weaker global asymptotic stability (GAS) property is established for a PI-PBC and the derivative action was not included.
\item[C2)] We show that for a specific class of power converters---and under some suitable conditions---there exists a GES equilibrium point for the closed-loop system even in the case where the desired operating point does not belong to set of assignable equilibria. We further establish an analytic relationship between such equilibrium and the desired operating point,  which is independent from the controller gains. Consistently with previously observed results, it is shown that small perturbations of the system parameters may generate large deviations from the desired operating point.
\item[C3)] Based on our recent work~\cite{zonetti2020}, we propose a modification of the PID-PBC by introducing a \textit{leakage} in the integral channel. We show that with this modification the stability properties of the PID-PBC are preserved, while increasing its performance and robustness, and with deviations from the desired operating point that can be adjusted by suitable design of the controller gains---similar to traditional primary controllers~\cite{vrana,simpson}. This analogy is further highlighted by the steady-state equations of the controller, which unveil a \textit{droop} characteristic between the control input and the passive output.
\item[C4)] We prove that an appropriate monotone transformation of the controller does not compromise the overall stability of the closed-loop system. A straightforward consequence of this result is that both the PID- and the P leaky ID (PLID)-PBC can be made robust to saturation of the control input.
\end{itemize}

 The paper is structured as follows. The model and a preliminary equilibria analysis are presented in Section \ref{sec:model}. Then, in Section~\ref{sec:GES}, we consider the system in closed-loop with the conventional PID-PBC and prove that under the assumption of perfect knowledge of the parameters, the system admits a GES equilibrium point for any positive gains. Robustness properties of the PID-PBC, for a specific class of power converters, are then analyzed in Section~\ref{sec:robustness}. To overcome some of the limitations of the PID-PBC, in Section~\ref{sec:leakyPI} we propose to introduce a leakage in the integral channel and analyze robust stability and performance properties of the resulting closed-loop system. Then, in Section~\ref{sec:saturation}, we further modify our design to guarantee that the control input is maintained within specific bounds, while preserving the stability. Finally, in Section~\ref{sec:applications}, we contextualize our findings to two widely diffused power applications: a boost dc/dc converter interfaced to a constant impedance, constant current (ZI) load and an HVDC grid-connected voltage source converter, further providing simulations that validate the aforementioned theoretical results. The paper is wrapped-up with some conclusions and guidelines for future works in Section \ref{sec:conclusions}.\\
 
\noindent \textbf{Notation.} All vectors are column vectors. Given positive integers $n$, $m$, the  symbols $0_{n}\in\mathbb{R}^n$ denotes the vector of all zeros, $0_{n\times m}\in\mathbb{R}^{n\times m}$ denotes the $n\times m$ column matrix of all zeros. Whenever clear from the context the latter is simply referred as $\underline 0\in\mathbb{R}^{n\times m}$. The symbols $\mathsf{1}_{n}\in\mathbb{R}^n$ denotes the vector of all ones, $\mathbb{I}_{n}\in\mathbb{R}^{n\times n}$ denotes the $n\times n$ identity matrix. Let $x=\mathrm{col}(x_1,\dots,x_n)\in\mathbb{R}^n$ a column vector with entries $x_i\in\mathbb{R}$.  Whenever clear from the context, we simply write $x=\mathrm{col}(x_i)$. Define the sets $\mathbb{R}^n_{\geq 0}:=\{x\in\mathbb{R}^n:\;x_i\geq 0\}$, $\mathbb{R}^n_{>0}:=\{x\in\mathbb{R}^n:\;x_i> 0\}$. Let $A=\mathrm{diag}(a_i)\in\mathbb{R}^{n\times n}$ a diagonal matrix with entries the scalars $a_i$ and $B=\mathrm{bdiag}(B_i)$ a block-diagonal matrix with entries the matrices $B_i$. For a function $f:\mathbb{R}^n\rightarrow\mathbb{R}$ the symbol $\nabla f$ denotes the transpose of its gradient. For a function $g:\mathbb{R}^n\rightarrow\mathbb{R}^m$ the term $\frac{\partial g(x)}{\partial x}\bigr\rvert_{\bar x}$ denotes the Jacobian of such a function evaluated at the point $\bar x\in\mathbb{R}^n$.    
  
  \section{Modelling of power converters \& preliminary analysis}\label{sec:model}
We consider the general class of power converters with no switching sources, and make the common assumption of the lossless high-frequency operation of the internal switches. As a result, switching dynamics can be safely neglected so that is possible to replace the power converter \textit{switched model} with a continuous-time \textit{averaged model}. For more information about this standard model reduction procedure and the limitations of the averaged model with respect to the switched model, the interested reader is referred to standard textbooks in power electronics~\cite{erickson2007fundamentals}. Based on Kirchhoff's laws the eletrical circuit of power converters with no switching sources can be described by bilinear systems of the form:
\begin{equation}
\begin{bmatrix}
L & \underline 0\\
\underline 0 & C
\end{bmatrix}
\begin{bmatrix}
\dot{i}_L\\
\dot{v}_C
\end{bmatrix}=-\begin{bmatrix}
R & \underline 0\\
\underline 0 & G
\end{bmatrix}
\begin{bmatrix}
{i}_L\\
{v}_C
\end{bmatrix}
+
\mathcal{J}(u)\begin{bmatrix}
{i}_L\\
{v}_C
\end{bmatrix}+
\begin{bmatrix}
v_0\\i_0
\end{bmatrix},
\end{equation}
with state vector $z:=\mathrm{col}(i_L,v_C)\in\mathbb{R}^{n}$, denoting the collection of currents flowing through the inductors $i_L\in\mathbb{R}^p$ and voltages across capacitors $v_C\in\mathbb{R}^q_{>0}$, where $n=p+q$; control vector $u:=\mathrm{col}(u_i)\in\mathbb{R}^m$, with $m<n$, denoting the collection of modulation indices; source vector $E:=\mathrm{col}(v_0,i_0)\in\mathbb{R}^n$, denoting the collection of energy sources, \textit{i.e.} the voltage and current sources $v_0\in\mathbb{R}_{>0}^p$ and $i_0\in\mathbb{R}^q$. We further define the inertia, dissipation and interconnection matrices 
$$
M:=\mathrm{bdiag}(L,C)\in\mathbb{R}^{n\times n},\quad \mathcal R:=\mathrm{bdiag}(R,G)\in\mathbb{R}^{n\times n},\quad 
\mathcal{J}(u):=\mathcal{J}_0+\sum_{i=1}^m\mathcal{J}_iu_i\in\mathbb{R}^{n\times n},
$$ 
where $L\in\mathbb{R}^{p\times p}$, $C\in\mathbb{R}^{q\times q}$,  $R\in\mathbb{R}^{p\times p}$ and $G\in\mathbb{R}^{q\times q}$ are diagonal positive definite matrices with entries given respectively by the inductances,  capacitances, resistances and conductances of the converter circuit, while $\mathcal{J}_i=-\mathcal{J}_i^\top\in\mathbb{R}^{n\times n}$, $i\in\{0\dots m\}$, are matrices characterizing the (possibly modulated) interconnection between the converter electrical components. Now, in order to derive a more compact representation of the converter---that further highlights the energy properties of the underlying electrical circuit---we find convenient to introduce the change of variables $x=Mz$, and consider the Hamiltonian function $\mathcal{H}:\mathbb{R}^n\rightarrow\mathbb{R}_{\geq 0}$:
\begin{equation}\label{eq:hamiltonian}
    \mathcal H(x):=\frac{1}{2}x^\top Qx,
\end{equation}
with $Q:= M^{-1}$, which represents the energy stored in the converter reactive components. Based on this definition, we can obtain the following energy-based description of the power converter, referred as\textit{ port-Hamiltonian} representation~\cite{escobar1999hamiltonian,GEOPLEX}:
\begin{equation}\label{eq:sys}
    \dot x=(\mathcal J_0+\sum_{i=1}^m\mathcal J_iu_i-\mathcal R)\nabla \mathcal H(x)+E,
\end{equation}
with the new state vector $x:=\mathrm{col}(\phi_L,q_C)\in\mathbb{R}^{n}$, denoting the collection of energy storing variables, \textit{i.e.} the fluxes of inductors $\phi_L\in\mathbb{R}^p$ and the charges of capacitors $q_C\in\mathbb{R}^q_{>0}$.
Note that the system \eqref{eq:sys} can be equivalently rewritten in the conventional input-affine form:
\begin{equation}
    \dot x=f(x)+g(x)u,
\end{equation}
 with vector field $f:\mathbb{R}^n\rightarrow\mathbb{R}^n$ and input matrix $g:\mathbb{R}^m\rightarrow\mathbb{R}^n$ given by:
\begin{equation}
\begin{aligned}
    f(x):=(\mathcal{J}_0-\mathcal{R})\nabla\mathcal{H}+E,\qquad g(x):=\begin{bmatrix}\mathcal J_1\nabla \mathcal H&\mathcal J_2\nabla \mathcal H&\dots \mathcal J_m\nabla \mathcal H\end{bmatrix}.
    \end{aligned}
\end{equation}
Because of the broad class of power converters described by \eqref{eq:sys} and the large variety of applications, control objectives can be of a very different kind. However, fundamental requirements, such as the ability to regulate the state near to suitable constant values, minimum performance, robustness to parameters uncertainties under limited sensing and actuation  must be achieved independently from the specific application. In order to characterize the steady-state conditions achievable via an appropriate control of the power converter, we introduce the notions of set of \textit{assignable equilibria} and \textit{equilibrium control} for the system  \eqref{eq:sys}. The set of assignable equilibria is given by the set
\begin{equation}\label{eq:equilibriaf}
\mathcal{E}:=\{x\in\mathbb{R}^n,\exists u\in\mathbb{R}^m:\;f(x,u)= 0_n\}\subset\mathbb{R}^n
\end{equation}

Note that if a full-rank left annihilator $g^\perp:\mathbb{R}^n\rightarrow\mathbb{R}^{(n-m)\times n}$ of $g$ exists, such a set can be written in compact form as:
\begin{equation}\label{eq:assignable}
    \mathcal E=\{ x\in\mathbb{R}^n:\;g^\perp( x)f( x)=  0_{n-m}\}.
\end{equation}
Let us denote $g^+:=(g^\top g)^{-1}g^\top$ as the Moore-Penrose left pseudoinverse of $g$. Then, for a given assignable equilibrium $\bar x\in\mathcal{E}$, the corresponding equilibrium control is given by:\footnote{In the remainder of the paper, for any $r$-dimensional signal $s:\mathbb{R}_{\geq 0}\rightarrow\mathbb{R}^r$ we denote as $\bar{s}$ the instance of such a signal at equilibrium conditions.} 
\begin{equation}\label{eq:eqcontrol}
\bar u:=\mathsf{u}(\bar x)=-g^+ (\bar{x})f(\bar x),
\end{equation}
 where the map $\mathsf{u}:\mathbb{R}^n\rightarrow \mathbb{R}^m$ is univocally defined. It is further possible to show that for a given constant control input $\bar u\in\mathbb{R}^m$, an equilibrium $\bar x\in\mathbb{R}^n$ exists and is always unique. To see this, consider the equilibria equations stemming from \eqref{eq:equilibriaf}, which are given by:
\begin{equation}\label{eq:equilibriacomment}
f(\bar x,\bar u)=(\mathcal{J}(\bar u)-\mathcal R)Q\bar x+E=  0_n.    
\end{equation}
Since $Q$ is positive definite, the algebraic equations \eqref{eq:equilibriacomment} admit a unique solution if and only if $\mathcal{J}(\bar u)-\mathcal R$ is nonsingular. Suppose now that $(\mathcal{J}(\bar u)-\mathcal R)v = 0_n$ for some $v\in\mathbb{R}^n$. Then $v^\top\left( \mathcal J(\bar u)-\mathcal R\right )v=-v^\top\mathcal R v=0$, which gives $\mathcal Rv =   0_n.$ Hence, since $\mathcal R$ is positive definite we necessarily have $v= 0_n$, thus implying that $\mathcal{J}(\bar u)-\mathcal R$ is always nonsingular and that the unique equilibrium point is given by:
\begin{equation}\label{eq:equilibrium-point}
\bar x=-Q^{-1}(\mathcal{J}(\bar u)-\mathcal R)^{-1}E.    
\end{equation}
 We now recall a fundamental result obtained in~\cite{hernandez2009adaptive} Proposition 1, the proof of which is omitted for brevity.\smallbreak

\begin{prop}[Passivity of the incremental model]\label{propassivity} Consider the system \eqref{eq:sys} and let $\bar x\in\mathcal{E}$, with $\bar u$ the corresponding equilibrium control. Then the map $\tilde u\rightarrow \tilde y$, with
\begin{equation}\label{eq:passiveoutput}
    \tilde u:=u-\bar u,\quad \tilde y:=g^\top (\bar x)Q \tilde x,
\end{equation}
is passive with storage function ${\mathcal{H}}(\tilde x)=\frac{1}{2}\tilde x^\top Q \tilde x$, where $\tilde x:=x-\bar x.$ In particular, the following power balance is verified:
\begin{equation}\label{eq:dotHpassivity}
    \dot{\mathcal{H}}(\tilde x)=-\tilde x^\top Q\mathcal{R}Q\tilde x+\tilde y^\top \tilde u.
\end{equation}
\end{prop}
\smallbreak
A straightforward consequence of Proposition~\ref{propassivity} is that whenever the system \eqref{eq:sys} is driven by a constant control input $\bar u$, the point $\bar x\in\mathcal E$ given by \eqref{eq:equilibrium-point} is a globally exponentially stable equilibrium of the controlled system. This unsurprising result stems from the fact that the system \eqref{eq:sys} is indeed a dissipative RLC circuit fed by constant current and voltage sources, and with zero energy net exchange through its control port due to the switching characteristic of the input. This aspect is neatly captured by the power balance:
\begin{equation}
    \underbrace{\dot{\mathcal{H}}(x)}_{\mathrm{stored}\;\mathrm{power}}=-\underbrace{x^\top Q\mathcal{R}Qx}_{\mathrm{dissipated}\;\mathrm{power}}+\underbrace{x^\top Qg(x)u}_{\mathrm{control}\;\mathrm{power}}+\underbrace{x^\top QE}_{\mathrm{supplied}\;\mathrm{power}},
\end{equation}
with the power supplied via the control port $x^\top Qg(x)u=\sum_ix^\top Q\mathcal{J}_iu_iQx$  equal to zero due to the skew-symmetry of $\mathcal{J}_i$.

\begin{rem}\textbf{(Alternative passive outputs).} Alternative passive outputs can be established using Lyapunov functions different than $\mathcal H(\tilde x)$---see~\cite{zhang2017pid} for a full characterization for port-Hamiltonian systems. A particular class is given by the outputs $\tilde y_2\in\mathbb{R}^m$ generated via quadratic functions $\mathcal{H}_2(\tilde x)=\frac{1}{2}\tilde{x}^\top P\tilde x$, where $P=P^\top\in\mathbb{R}^{n\times n}$ is positive definite. Indeed, by calculating its derivative along the system's trajectories we obtain:
\begin{equation}
\dot{\mathcal{H}_2}=\tilde{x}^\top P\dot{\tilde x}=\tilde{x}^\top \left(\frac{P F(\bar u) + F^\top (\bar u)P}{2}\right)\tilde x+\tilde y_2^\top \tilde u,\qquad F(\bar u):=(\mathcal{J}_0+\sum\mathcal{J}_i\bar u_i-\mathcal{R})Q,
\end{equation}
with output $\tilde y_2:=g^\top (x)P\tilde x.$  Unless $P=\epsilon Q$, with $\epsilon>0$, the output is a nonlinear function of  the converter variables with no obvious physical interpretation and the passivity of which is subject to the verification of the equilibrium-dependent Lyapunov equation:
\begin{equation}\label{eq:Lyapunov-eq}
P F(\bar u)+F^\top(\bar u)P<0,
\end{equation}
a fact that complicates its use as driving signal for conventional PID controllers, and that considerably reduces its appeal for industrial applications. Instead, with $P=\epsilon Q$, the Lyapunov equation \eqref{eq:Lyapunov-eq} is always verified, the passive output $\tilde y_2=\epsilon\tilde y$ is linear and generated via the incremental version of the natural energy function---reasons for which it retained a lot of attention from practitioners. A further reason for its popularity is the striking similarity with the power output employed by the celebrated Akagi's controller~\cite{akagi2017}. Based on these considerations, in the remainder of the paper we will focus explicitly on the passive output $\tilde y$, leaving the analysis and design of controllers based on alternative passive outputs for future investigation.
\end{rem}

\begin{rem}\textbf{(Performance limitations).}\label{rem:minper}
Albeit stability is preserved, because of the converter circuit small losses the open-loop control action in general fails to properly damp the oscillations generated by the converter reactive components. Poor performances are then exhibited by the open-loop controlled system, since the convergence rate of the trajectories is bounded by the time constant of the dominant electrical component. The worst-case performances guaranteed by  suitably defined PID-PBCs are investigated in Section~\ref{sec:GES} and Section~\ref{sec:leakyPI}.
\end{rem}

\begin{rem}\textbf{(Parameters uncertainty).}\label{rem:estimcalE}
The set of assignable equilibria $\mathcal{E}$ is defined via the vector field $f$ and the input matrix $g$, for which only an estimation may be available for the control design. The implications of such imprecise knowledge on the robustness of the system in closed-loop with PID-PBCs are thoroughly discussed in Section~\ref{sec:robustness} and Section~\ref{sec:leakyPI}.
\end{rem}

\begin{rem}\textbf{(Actuation limitations).}\label{rem:boundU}
As it will be made clear later in the paper, the modulation indices are actually constrained to a closed set $\mathcal{U}\subset\mathbb{R}^m$.  Nevertheless, for clarity of presentation we find convenient to provide our results first for the model with unconstrained control input. The validity of such results with bounded control input is addressed in detail in Section~\ref{sec:saturation}.
\end{rem}

\begin{rem}\textbf{(Sensing limitations).}
In this paper we assume that all state variables can be measured and thus are readily available for control design. While this is not always the case in practical applications, this assumption can be justified by the design of a finite-time convergence observer along the lines of~\cite{bobtsov2020globally} and for which the results of this paper can be used \textit{mutatis mutandi}.
\end{rem}

\section{Global exponential stability of the PID-PBC}\label{sec:GES}
Motivated by the poor performances of the open-loop controlled system---see Remark~\ref{rem:minper}---we find convenient to introduce a feedback control action---the most popular form being a PID controller. Let then $x_\star\in\mathbb{R}^n$ be a given reference vector and define the PID controller:\footnote{In the remainder of the paper, for any $r$-dimensional signal $s:\mathbb{R}_{\geq 0}\rightarrow\mathbb{R}^r$ we denote $ s_\star$ as a given constant reference for such a signal.}
 \begin{align}
    \dot x_c&=-g^\top (x_\star)Qx,\label{eq:PIPBC-I}\\
    u&=-K_Pg^\top (x_\star)Qx+K_Ix_c-K_Dg^\top (x_\star)Q\dot x\label{eq:PIPBC-P}.
\end{align}
where $K_P\in\mathbb{R}^{m\times m}$, $K_I\in\mathbb{R}^{m\times m}$ and $K_D\in\mathbb{R}^{m\times m}$ are gain matrices. Note that whenever $x_\star\in\mathcal{E}$, the resulting PID controller is driven by the passive output \eqref{eq:passiveoutput} and for this reason is commonly referred as a \textit{PID-PBC}~\cite{hernandez2009adaptive}. With a little abuse of the terminology, and recalling that the controller is rooted on the same passivity concept, we will refer to \eqref{eq:PIPBC-I}-\eqref{eq:PIPBC-P}  as a PID-PBC even if $x_\star\notin\mathcal{E}$.  We have then the following proposition.\smallbreak

\begin{prop}[GES of the PID-PBC]\label{prop:GES}
Consider the system \eqref{eq:sys} in closed-loop with the controller \eqref{eq:PIPBC-I}-\eqref{eq:PIPBC-P}. Let $x_\star\in\mathcal{E}$, with $u_\star:=\mathsf{u}(x_\star)$ the corresponding equilibrium control. Then the point $(x_\star,K_I^{-1}u_\star)\in\mathcal{E}\times\mathbb{R}^m$ is a globally exponentially stable equilibrium of the closed-loop system for any positive semidefinite gain matrices $K_P,K_D$ and positive definite gain matrix $K_I$.
\end{prop}
\begin{pf} 
The equilibria of \eqref{eq:sys} in closed-loop with \eqref{eq:PIPBC-I}-\eqref{eq:PIPBC-P} are the solution $(\bar x,\bar x_c)\in\mathcal{E}\times \mathbb{R}^m$ of the following equations:
\begin{equation}\label{eq:equilibriasys}
\begin{aligned}
0_n&=(\mathcal J_0-\mathcal{R})Q\bar x+g(\bar x)\bar u +E\\
0_m&=g^\top(x_\star)Q\bar x,
\end{aligned}
\end{equation}
with $\bar u:=K_I\bar x_c$. Since $x_\star\in\mathcal{E}$, \eqref{eq:equilibriasys} hold true for $\bar x=x_\star$ and $\bar x_c=K_I^{-1}u_\star$. Hence, $(x_\star,K_I^{-1}u_\star)$ is an equilibrium of the closed-loop system.  Define the incremental variables:
$$
\tilde x=x-\bar x,\quad \tilde x_c=x_c-\bar x_c,\quad \tilde u=u-K_I\bar x_c,
$$
and, to simplify the notation, let $g(\bar x):=\bar g$. Using \eqref{eq:equilibriasys} and recalling that $\bar g^\top Q\bar x=0$, we can obtain the incremental model for \eqref{eq:sys} in closed-loop with \eqref{eq:PIPBC-I}-\eqref{eq:PIPBC-P},  that is:
\begin{align}
\dot x&=(\mathcal J_0-\mathcal{R})Q\tilde x+g(\tilde x)(\tilde u+\bar u)+\bar g \tilde u \label{eq:incsys}\\
\dot x_c&=-\bar g^\top Q\tilde x.\label{eq:incI-PBC}
\end{align}
with
\begin{equation}\label{eq:incP-PBC}
    \tilde u=-K_P\bar g^\top Q\tilde x+K_I\tilde x_c-K_D\bar g^\top Q\dot x.
\end{equation}
Now define the following energy function 
\begin{equation}\label{eq:Veps}
    \mathcal{V}_\varepsilon(\tilde x,\tilde{x}_c):=\mathcal{H}(\tilde x)+\frac{ 1}{2}\tilde{x}^\top Q\bar gK_D\bar g ^\top Q\tilde x+\frac{ 1}{2} \tilde x_c^\top K_I\tilde x_c- \varepsilon \tilde{x}^\top Q \bar g K_I\tilde x_c=\dfrac{1}{2}
    \begin{bmatrix}
    Q\tilde x \\K_I \tilde x_c
    \end{bmatrix}^\top 
    \underbrace{\begin{bmatrix}
    Q^{-1}+\bar gK_D\bar g^\top & -\varepsilon  \bar g \\
    -\varepsilon  \bar g^\top & K^{-1}_I
    \end{bmatrix}}_{\mathcal{Q}_\varepsilon}
     \begin{bmatrix}
    Q\tilde x \\K_I\tilde x_c
    \end{bmatrix},
\end{equation}
with $\varepsilon\geq 0$. Then $\mathcal{V}_\varepsilon$ is positive definite if and only if $\mathcal Q_\varepsilon>0$ and, if this is the case, it is also radially unbounded. Since $Q>0$, $K_I>0$ and $K_D\geq 0$, this is equivalent to:
\begin{equation}\label{eq:positiveW}
Q^{\frac{1}{2}}\bar g(\varepsilon ^2K_I-K_D)\bar g^\top Q^{\frac{1}{2}} <\mathbb{I}_n.
\end{equation}
Let us define $\mathcal{V}_0$ as the energy function \eqref{eq:Veps} when we take $\varepsilon=0$. Note from \eqref{eq:positiveW} that such a function is always positive definite being $K_D\geq 0$. We get then:
\begin{equation}\label{eq:dotV0}
\begin{aligned}
    \dot{\mathcal{V}}_0&=\dot{\mathcal{H}}+\tilde{x}^\top Q\bar gK_D \bar g^\top Q\dot x+\tilde{x}_c^\top K_I\dot{\tilde{x}}_c\\
&=-\tilde{x}^\top Q\mathcal{R}Q\tilde x+ \tilde{x}^\top Q\bar g \left(-K_P \bar g^\top Q\tilde x+ K_I\tilde x_c-K_D \bar g^\top Q\dot x\right)+\tilde{x}^\top Q\bar gK_D\bar g ^\top Q\dot x+  \tilde x_c^\top K_I \left(-\bar g^\top Q \tilde x\right)\\
&=-\tilde{x}^\top Q(\mathcal{R}+\bar gK_P\bar g^\top) Q\tilde x.
    \end{aligned}
\end{equation}
where the second equivalence is obtained using \eqref{eq:dotHpassivity} and by substitution of \eqref{eq:incI-PBC} and \eqref{eq:incP-PBC}. Since $K_P\geq 0,\mathcal{R}>0$, we have $\dot{\mathcal{V}}_0\leq 0$, from which follows, by Lyapunov's arguments, that the trajectories of the system are bounded, and so is $u$. To prove global exponential stability, we define:
\begin{equation}
A:= Q(\mathbb{I}_n+\bar gK_D\bar g^\top Q)\in\mathbb{R}^{n\times n},\qquad b(u):= \mathcal{J}_0+\sum_{i=1}^m\mathcal{J}_iu_i-\mathcal{R}-\bar gK_P\bar g^\top.
\end{equation}
 Hence, from \eqref{eq:incP-PBC} and \eqref{eq:incsys} we obtain 
\begin{equation}
    (\mathbb{I}_m+\bar gK_D\bar g^\top Q)\dot x=b(u)Q\tilde x+\bar g K_I\tilde x_c,
    \end{equation}
    which implies
    \begin{equation}\label{eq:dotxb}
        \dot x=A^{-1}Q\left[b(u)Q\tilde x+\bar g K_I\tilde x_c\right].
    \end{equation}
Note that $A$ is the sum of a positive definite and a positive semidefinite matrix and is thus invertible. Now consider $\mathcal{V}_\varepsilon$, where $\varepsilon$ is a positive constant to be later determined. The derivative of $\mathcal{V}_\varepsilon$ along the system's trajectories is then given by:
\begin{equation}\label{eq:dotW}
\begin{aligned}
    \dot{\mathcal{V}}_\varepsilon&=\dot{\mathcal{V}}_0-\varepsilon\tilde{x}^\top Q\bar gK_I\dot{\tilde x}_c-\varepsilon\tilde{x}_c^\top K_I\bar g^\top Q\dot{\tilde x}\\
    &=-\tilde{x}^\top Q(\mathcal{R}+\bar gK_P\bar g^\top) Q\tilde x- \varepsilon \tilde x^\top Q\bar g K_I \left( - \bar g^\top Q\tilde x\right)- \varepsilon \tilde{x}_c^\top K_I\bar g^\top QA^{-1}Q\left[b(u)Q\tilde x+\bar g K_I\tilde x_c\right]\\
     &=-\tilde{x}^\top Q \left[\mathcal{R}+\bar g\left( K_P -\varepsilon K_I\right)\bar g^\top\right]Q\tilde x- \tilde x_c^\top K_I( \varepsilon\bar g^\top QA^{-1}Q \bar g) K_I\tilde x_c-\tilde x^\top Q\left[ \varepsilon  b^\top (u) QA^{-\top}Q \bar g\right]K_I\tilde x_c,
\end{aligned}    
\end{equation}
where the second equivalence follows from \eqref{eq:dotV0}, \eqref{eq:incI-PBC} and \eqref{eq:dotxb}. Now let us define the positive definite matrix $B:=QA^{-1}Q\in\mathbb{R}^{n\times n}$, so we can rewrite \eqref{eq:dotW} in compact form: 
\begin{equation}\label{eq:dotwP}
    \dot{\mathcal{V}}_\varepsilon=-\begin{bmatrix} Q \tilde x\\K_I\tilde x_c
    \end{bmatrix}^\top
    \underbrace{\begin{bmatrix}
    \mathcal{R}+\bar g( K_P-\varepsilon K_I)\bar g^\top & \frac{1}{2}\varepsilon b^\top B^\top\bar g\\
    \frac{1}{2} \varepsilon\bar g^\top B b & \varepsilon\bar g^\top B\bar g
    \end{bmatrix}}_{\mathcal D_\varepsilon}\begin{bmatrix} Q\tilde x\\K_I\tilde x_c
    \end{bmatrix},
\end{equation}
which is strictly negative if and only if $\mathcal D_\varepsilon>0$, that is:
\begin{equation}\label{eq:negativedotW}
\mathcal{R}+\bar gK_P\bar g^\top >  \varepsilon\left[\bar g K_I\bar g^\top+\dfrac{1}{4}  b^\top B^\top\bar g(\bar g^\top B\bar g)^{-1}\bar g^\top B b\right].
\end{equation}
\noindent 
Since $u$ is bounded, the term $\mathcal{B}:=b^\top B^\top\bar g(\bar g^\top B\bar g)^{-1}\bar g^\top B b$ is bounded as well, and is immediate to conclude that it is always possible to pick a sufficiently small $\varepsilon$ verifying both \eqref{eq:positiveW} and \eqref{eq:negativedotW}, thus completing the proof.
\end{pf}
\begin{rem}\textbf{(Performance limitations of the PID-PBC).} The global exponential stability result of Proposition \ref{prop:GES} implies that for any $\varepsilon>0$ verifying \eqref{eq:positiveW} and \eqref{eq:negativedotW}, there exists an $\alpha_\varepsilon > 0$ such that 
\begin{equation}\label{eq:rate}
\dot{\mathcal{V}}_\varepsilon \leq -\alpha_\varepsilon \cdot \mathcal{V}_\varepsilon(\tilde{x},\tilde{x}_c),\qquad
\alpha_\varepsilon := 2 \dfrac{\lambda_\mathrm{m}(\mathcal D_\varepsilon)}{\lambda_\mathrm{M}(\mathcal Q_\varepsilon)},
\end{equation}
\noindent with $\max(\alpha_\varepsilon)$ representing an exponential convergence rate bounding the trajectories of the system, and where  $\lambda_\mathrm{m}$, $\lambda_\mathrm{M}$ denote the minimum and maximum eigenvalues of the corresponding matrix, respectively. Performances of the PID-PBC are also expected to be highly sensitive to measurements noise, as this is unavoidably amplified via the computation of the derivative of voltages and currents. This problem can be obviated by omission of the derivative action or mitigated by appropriate filtering of such signals~\cite{isaksson2002derivative}.
\end{rem}

\begin{rem}\textbf{(Design of the Lyapunov function).} The design of a parametrized cross-term in the function~\eqref{eq:Veps}   is inspired by~\cite{weitenberg2018exponential} and is motivated by the lack of damping in the controller state variables---a fact that stymies the construction of a \textit{strict} Lyapunov function for the closed-loop system. Indeed, if $\varepsilon=0$ the function \eqref{eq:Veps} coincides with the \textit{non-strict} Lyapunov function employed in~\cite{hernandez2009adaptive} which was used to prove asymptotic, but not exponential, convergence of the system's trajectories.
\end{rem}

\begin{rem}\textbf{(Networked system scalability).} Although the control design is analyzed and developed for the case of a single power converter,  it is possible to show that an identical, \textit{decentralized} solution can be obtained for a power system constituted by an arbitrary number of power converters interfaced via a linear DC network. Hence all results reported in this paper applies \textit{mutatis mutandi} to the interconnected case. This property, which was exploited in~\cite{Zonetti2015} for the case of HVDC transmission systems, stems from the fact that the feedback interconnection of port-Hamiltonian systems of the form \eqref{eq:sys} results in a system of the same form.   
\end{rem}

 \section{Robustness margins}\label{sec:robustness}
From Proposition \ref{prop:GES} we deduce that an underlying pre-requisite for the design of the PID-PBC is that the reference vector $x_\star$ belongs to the set of assignable equilibria $\mathcal{E}$. Such a vector is typically determined by an higher-level references calculator that takes as input a vector $x_\mathrm{1d}\in\mathbb{R}^m$ of desired values and computes via \eqref{eq:assignable} the vector $x_{2\star}\in\mathbb{R}^{n-m}$ of remaining components to guarantee that its output $x_\star=\mathrm{col}(x_\mathrm{1d},x_\mathrm{2\star})\in\mathcal{E}$. In this context, as suggested by Remark~\ref{rem:estimcalE}, the knowledge of the vector field $f$ and input matrix $g$ represents a critical issue. Indeed, in a practical scenario the uncertainty on the system's dissipation and supply/demand results in an approximate knowledge of the dissipation matrix $\mathcal{R}$ and sources vector $E$, which ultimately allow to establish only an \textit{estimated} set of assignable equilibria $\hat{\mathcal{E}}$ that does not coincide in general with the \textit{actual} set $\mathcal E$. As  in this case the reference calculator will unavoidably generate unassignable references vectors for the PID-PBC, it may be questioned whether the controller is robust to such uncertainty. \\
We now restrict our attention to a class of power converters described by \eqref{eq:sys} that further verify the following assumption.\smallbreak

 \begin{assum}\label{ass:rank} The rank of the matrix $g(x)$ is equal to $n-1$.
\end{assum}

This assumption is verified for all power converters with an electrical scheme constituted by $n-1$ half-bridges, that share at least a reactive component~\cite{erickson2007fundamentals}. This is a common architecture in many power electronic converters, and such a broad class includes, among the others, the popular boost and two-level voltage source converters, which are currently the most diffused topologies for grid applications.  \\
An immediate consequence of this assumption is that a full-rank left annihilator of the input matrix is given by $g^\perp(x)=x^\top Q$ and then the set of assignable equilibria \eqref{eq:assignable} can be rewritten as:
\begin{equation}\label{eq:powerflow}
\mathcal{E}=\{x\in\mathbb{R}^n:-x^\top Q\mathcal{R}Qx+E^\top Qx= 0\}.
\end{equation}
For a given $\bar x\in\mathcal{E}$ the underlying equation is also referred as \textit{power flow equation}, with the terms
$$P_\mathrm{loss}(\bar x):=\bar x^\top Q\mathcal{R}Q\bar x ,\quad P_\mathrm{net}(\bar x):=E^\top Q\bar x$$  denoting respectively the dissipated power and net supplied  power at steady-state.
To analyze robustness of the PID-PBC we assume that the reference vector $x_\star$ belongs to a set of \textit{estimated} assignable equilibria $\hat{\mathcal{E}}\subset\mathbb{R}^{n}$ defined as follows:
\begin{equation}\label{eq:hatE}
    \hat{\mathcal{E}}:=\{x\in\mathbb{R}^n:\;-x^\top Q\hat{\mathcal{R}} Qx+\hat{E}^\top Qx= 0\},
\end{equation}
 where the underlying equations are referred as \textit{estimated power flow equations}, while $\hat{\mathcal{R}}=\hat{\mathcal{R}}^\top\in\mathbb{R}^{n\times n}$, with $\hat{\mathcal{R}}>0$ and $\hat E\in\mathbb{R}^n$ denote respectively the estimated dissipation matrix and the estimated source vector. Note that only if $\mathcal{R}=\hat{\mathcal{R}}$, $E=\hat E$, \textit{i.e.}, if we have exact knowledge of the system parameters, the actual power flow defined by  \eqref{eq:powerflow} and the estimated power flow coincide. We have then the following proposition.\smallbreak

 \begin{prop}[Robustness of the PID-PBC]\label{prop:gamma}
Consider the system \eqref{eq:sys} in closed-loop with the controller \eqref{eq:PIPBC-I}-\eqref{eq:PIPBC-P}, and assume it verifies Assumption \ref{ass:rank}. Let $x_\star\in\hat{\mathcal{E}}$,  $u_\star:=\mathsf{u}(\gamma x_\star)$, with
\begin{equation}\label{gamma}
\gamma:=\frac{P_\mathrm{net}(x_\star)}{P_\mathrm{loss}(x_\star)}.
\end{equation}
Then, if $P_\mathrm{net}(x_\star)>0$, the point $(\gamma x_\star,K_I^{-1}u_\star)\in\mathcal{E}\times\mathbb{R}^m$ is a GES equilibrium of the closed-loop system for any positive semidefinite gain matrices $K_P,K_D$ and positive definite gain matrix $K_I.$
\end{prop}
 \begin{pf} 
The equilibria of the closed-loop system are the solutions $(\bar x,\bar x_c)\in\mathcal{E}\times \mathbb{R}^m$ of the following equations:
\begin{align}
0_n&=(\mathcal J_0-\mathcal{R})Q\bar x+\bar g(\bar x)\bar u +E\label{eq:syseq}\\
0&=g^\top (x_\star) Q\bar x\label{eq:collinear},
\end{align}
with equilibrium control $\bar u:=\mathsf{u}(\bar x)=K_I\bar x_c$, from which we obtain immediately:
\begin{equation}\label{eq:barxc}
\bar x_c=K_I^{-1}\mathsf{u}(\bar x)=-K_I^{-1}g^+ (\bar x)\left[(\mathcal J_0-\mathcal R) Q\bar x+E\right].
\end{equation}
To simplify the notation, let us introduce the definitions $\bar g:=g(\bar x)$, $g_\star:=g(x_\star).$ From \eqref{eq:collinear} we deduce then that $Q\bar x\in\ker g_\star^\top$. On the other hand, since $g_\star^\top Qx_\star= 0$, we also have $Qx_\star\in\ker g_\star^\top$. Now, note that Assumption~\ref{ass:rank} implies that the dimension of $\ker g_\star^\top$ is equal to $1$. Hence, since both vectors $\bar x$ and $x_\star$ belong to such a set, they must be collinear, i.e. there exists a constant $\gamma\in\mathbb{R}$ such that $\bar x=\gamma x_\star.$\\
We next show that such $\gamma$ is indeed given by~\eqref{gamma}. Recall that $(\bar x,\bar x_c)$ is an equilibrium of the closed-loop system and then $\bar x$ must be necessarily assignable, i.e. $\bar x\in\mathcal{E}$. Hence, the following holds:
\begin{equation}\label{eq:pflow2}
-\bar x^\top Q\mathcal{R}Q\bar x+E^\top Q\bar x= 0,
\end{equation}
and then, by replacing $\bar x=\gamma x_\star$ therein, we get
\begin{equation}
-\gamma^2 x_\star^\top Q\mathcal{R}Qx_\star+\gamma E^\top Qx_\star= 0,
\end{equation}
from which it is easy to derive \eqref{gamma}. Finally, by replacing $\bar x=\gamma x_\star$ into \eqref{eq:barxc} we obtain $\bar x_c=K_I^{-1}\mathsf{u}(\gamma x_\star)$ and then have demonstrated that $(\gamma x_\star,K_I^{-1}u_\star)$ is an equilibrium of the closed-loop system. To prove global exponential stability, let us introduce the change of variable $\xi:=\gamma x_c$. Therefore, the controller \eqref{eq:PIPBC-I}-\eqref{eq:PIPBC-P} can be rewritten equivalently as:
\begin{align}
    \dot \xi&=-\bar g^\top Qx,\label{eq:pigamma}\\
    u&=-\kappa_P\bar g^\top Qx+\kappa_I\xi-\kappa_D\bar g^\top Q\dot x,\label{eq:pigamma2}
\end{align}
with $\kappa_P:=K_P/\gamma$, $\kappa_I:=K_I/\gamma$, $\kappa_D:=K_D/\gamma$. Finally, since $P_\mathrm{net}(x_\star)>0$, it follows that $\gamma>0$  and then $\kappa_P$, $\kappa_I$, $\kappa_D$ are positive (semi)definite if and only if $K_P$, $K_I$, $K_D$ are respecively positive (semi)definite. The proof is completed recalling that since $\bar x\in\mathcal E$, the controller \eqref{eq:pigamma}-\eqref{eq:pigamma2} verifies the assumptions of Proposition \ref{prop:GES}. 
\end{pf}
\begin{rem}\label{rem:deviations} \textbf{(Steady-state deviations under PID-PBC).} 
In case of poor knowledge of the system parameters, the PID-PBC is unable to regulate the $n$ components of the state to their precise reference values. Indeed, the normalized steady-state deviation of the actual equilibrium component $\bar x_i$ from the reference vector component $x_{i\star}$ is the same for any $i\in[1\;n]$ and it is given by the scalar $\Delta x\in\mathbb{R}_{\geq 0}$, with:
$$\Delta x:=\Bigg\vert\frac{ \bar x_i-x_{i\star}}{x_{i\star}}\Bigg\vert= \vert \gamma-1\vert.$$
Note however that in many practical applications, exact, \textit{e.g.} voltage, regulation is not a strict requirement and for a safe operation of the system it is enough to guarantee that the trajectories settle \textit{sufficiently close} to a desired value~\cite{vrana}.
\end{rem}

\begin{rem}\textbf{(Estimated passive output: zero dynamics).} Along the lines of the proof of Proposition~\ref{prop:gamma}, it is possible to show that, under Assumption~\ref{ass:rank}, the zero dynamics $\dot{\zeta}=f_0(\zeta)$, associated to \eqref{eq:sys} and the \textit{estimated} passive output $\hat y:=g(x_\star)^\top Qx$ is given by the following (stable) linear system:
\begin{equation}
    \mathcal{H}(x_\star)\dot \zeta=-P_\mathrm{loss}(x_\star)\zeta+P_\mathrm{net}(x_\star),
\end{equation}
which has an equilibrium at $\bar \zeta:=\gamma$. This result is a generalization of the result presented in~\cite{Zonetti2015} for the case of HVDC transmission systems.
\end{rem}

 \begin{rem} \textbf{(Estimated equilibrium feasibility).}
The term $P_\mathrm{net}(x_\star)$ corresponds to the net supplied power that the PID-PBC seeks to impose to the converter in order to achieve regulation to the reference vector $x_\star\in\hat{\mathcal{E}}$. As a result, the condition $P_\mathrm{net}(x_\star)>0$ merely states that such power must be feasible, \textit{i.e.}, positive. Note that this condition is independent from the controller parameters and from the definition of $\hat{\mathcal E}$.
\end{rem}

\begin{rem}\textbf{(Robustness of the PID-PBC for general power converters).} For power converters that do not verify Assumption~\ref{ass:rank} a robustness analysis can be still performed. Indeed, Proposition~\ref{prop:GES} suggests that the \textit{virtual} damping $\varepsilon$ introduced via the Lyapunov function $\mathcal{V}_\varepsilon$ can be used to dominate, in the function's derivative, the additional term resulting from the mismatch $x_\star-\bar x$ between the estimated and actual equilibria. Conditions for the convergence of the system's trajectories to a constant steady-state can be thus established. A similar approach is developed in Section~\ref{sec:leakyPI}, where an \textit{actual} damping is introduced in the integral channel.
\end{rem}

\section{Droop design via P leaky ID-PBC}\label{sec:leakyPI}
In the previous sections it was observed that the PID-PBC possesses important global stabilization and robustness properties. However, several limitations exist, especially for lightly damped systems. First, the controller may exhibit poor performances, a fact that can be explained by the possibly slow convergence rate given by \eqref{eq:rate}. Second, 
the stability condition $P_\mathrm{net}(x_\star)>0$ established in Proposition~\ref{prop:gamma} eventually determines narrow robustness margins, which may result in loss of stability in presence of large variations of the source vector $E$. Third, even if stability is preserved, small deviations of the reference vector $x_\star$ from the set of assignable equilibria $\mathcal E$ may induce closed-loop equilibria that are located far away from operating points that are of physical interest---independently from the controller gains.  A more in-depth analysis of these issues is postponed to Section~\ref{sec:applications}, where specific power applications are investigated.\\
To overcome the aforementioned limitations, we propose to introduce a leakage in the integral channel of the PID-PBC \eqref{eq:PIPBC-I}-\eqref{eq:PIPBC-P}, so that the new controller, in the sequel referred as P leaky ID (PLID)-PBC, is given by:
 \begin{align}
    \dot x_c&=-g^\top (x_\star)Qx-K_L K_I(x_c-x_{c\star}),\label{eq:PLIPBC-I}\\
    u&=-K_Pg^\top (x_\star)Qx+K_Ix_c-K_Dg^\top(x_\star) Q\dot x\label{eq:PLIPBC-P}.
\end{align}
where $x_\star\in\hat{\mathcal{E}}$ and $x_{c\star}:=K_I^{-1}\mathsf{u}(x_\star)\in\mathbb{R}^m$, with the map $\mathsf{u}$ given by \eqref{eq:eqcontrol}, are suitably defined reference vectors, and $K_P\in\mathbb{R}^{m\times m}$, $K_I\in\mathbb{R}^{m\times m}$, $K_D\in\mathbb{R}^{m\times m}$, $K_L\in\mathbb{R}^{m\times m}$ are gain matrices. We have then the following proposition.\smallbreak 

\begin{prop}[GES of the PLID-PBC]\label{prop:PLI}
Consider the system \eqref{eq:sys} in closed-loop with the controller \eqref{eq:PLIPBC-I}-\eqref{eq:PLIPBC-P}. Assume that there exists an equilibrium point $(\bar x,\bar x_c)\in\mathcal E\times\mathbb{R}^m$ for the closed-loop system and let 
\begin{equation}\label{eq:K}
    \mathcal{K}_P:=\frac{1}{2}\left[g(\bar x)K_Pg^\top (x_\star)+g( x_\star)K_Pg^\top (\bar x)\right],\qquad \mathcal{K}_D:=\frac{1}{2}\left[g(\bar x)K_Dg^\top (x_\star)+g( x_\star)K_Dg^\top (\bar x)\right]
\end{equation}
Then if the following inequalities hold:
\begin{equation}\label{eq:condition}
   \mathcal R+ \mathcal K_P>0,\quad Q^{-1}+ \mathcal K_D>0,\quad K_L >\frac{1}{4}g^\top (x_\star-\bar x)(\mathcal{R}+\mathcal{K}_P)^{-1}g(x_\star-\bar x),
\end{equation}
the equilibrium $(\bar x,\bar x_c)$ is GES  for any positive definite gain matrix $K_I$. Moreover, if $x_\star\in\mathcal{E}$, then $(x_\star,x_{c\star})\in\mathcal E\times\mathbb{R}^m$ is a GES equilibrium for any positive semidefinite gain matrices $K_P,K_D$ and positive definite gain matrices $K_I,K_L.$
 \end{prop}
\begin{pf}
  Recalling that an equilibrium exists by assumption, it is a solution $(\bar x,\bar x_c)\in\mathcal E\times\mathbb{R}^m$ of the following equations:
\begin{align}
         0_n&=(\mathcal J_0- \mathcal R)Q\bar x+E+g(\bar x)\bar u\label{eq:equilibria1}\\
         0_m&=-g^\top (x_\star) Q\bar x-K_L K_I(\bar x_c-x_{c\star}),\label{eq:equilibria2}
 \end{align}
with:
\begin{equation}\label{eq:baru}
\bar u=-K_Pg^\top (x_\star) Q\bar x+K_I\bar x_c.
\end{equation}
To simplify the notation, let us define $\bar g:=g(\bar x)$ and $g_\star:=g(x_\star).$ Now consider the incremental variables
\begin{equation}
    \tilde u:=u-\bar u,\quad \tilde{x}:=x-\bar x,\quad \tilde{x}_c:=x_c-\bar x_c,
    \end{equation}
    and the incremental energy function 
    $$
        {\mathcal{V}_\star}(\tilde x,\tilde x_c):=\mathcal{H}(\tilde x)+\frac{ 1}{2}\tilde{x}^\top Q\bar gK_D g_\star^\top Q\tilde x+\frac{1}{2}\tilde x_c^\top K_I\tilde x_c=
        \frac{1}{2}\begin{bmatrix}
    Q\tilde x \\K_I \tilde x_c
    \end{bmatrix}^\top 
    \begin{bmatrix}
    Q^{-1}+\mathcal{K}_D & 0 \\
    0& K^{-1}_I
    \end{bmatrix}
     \begin{bmatrix}
    Q\tilde x \\K_I\tilde x_c
    \end{bmatrix},
        $$
        which is positive definite and radially unbounded, being $K_I>0$ and $Q^{-1}+\mathcal{K}_D>0$ by hypothesis.   Using \eqref{eq:PLIPBC-P}  and \eqref{eq:baru} we obtain:
    \begin{equation}\label{eq:tildeu} 
    \begin{aligned}
        \tilde u&=-K_Pg_\star^\top Q(\tilde x+\bar x)+K_I(\tilde x_c+\bar  x_c)+K_Pg_\star^\top Q\bar x-K_I\bar  x_c-K_Dg_\star^\top Q\dot x\\
        &=-K_Pg_\star^\top Q\tilde x+K_I\tilde x_c-K_Dg_\star^\top Q\dot x.
        \end{aligned} 
    \end{equation}
 On the other hand, from \eqref{eq:PLIPBC-I} we have:
\begin{equation}\label{eq:tildexi}
    \begin{aligned}
        \dot{\tilde{x}}_c=&-K_L K_I\tilde  x_c-K_L K_I\bar  x_c-g_\star^\top Q\tilde x-g_\star ^\top Q\bar  x +K_L K_Ix_{c\star}\\
        =&-K_L K_I\tilde  x_c-\bar g^\top  Q\tilde x-(g_\star-\bar g)^\top  Q\tilde x,
    \end{aligned}
\end{equation}
where in the last equivalence we used \eqref{eq:equilibria2} and added and subtracted the term $\bar g^\top Q\tilde x$. 
The derivative of $\mathcal{V}_\star$ along the trajectories of the system then reads:
\begin{equation} 
\begin{aligned}
    \dot{\mathcal{V}}_\star=&\;\dot{\mathcal{H}}+\tilde{x}^\top Q\bar g K_Dg_\star^\top Q\dot x+\frac{1}{2}\tilde x_c K_I\dot{\tilde x}_c\\
    =&-\tilde x^\top Q\mathcal{R}Q\tilde x+\tilde x^\top Q\bar g \tilde u+\tilde{x}^\top Q\bar g K_Dg_\star^\top Q\dot x-\tilde x_c^\top K_IK_L K_I\tilde  x_c-\tilde{x}_c^\top K_I\bar g^\top Q\tilde x-\tilde{x}_c^\top K_I(g_\star-\bar g)^\top Q\tilde{x}\\
    =&-\tilde x^\top Q\left(\mathcal{R}+\bar g K_P g_\star^\top \right)Q\tilde x+\tilde x^\top Q\bar g K_I\tilde  x_c-\tilde x_c^\top K_IK_L K_I\tilde  x_c-\tilde{x}_c^\top K_I(g_\star-\bar g)^\top Q\tilde{x}
    \end{aligned} 
    \end{equation}
where the second equivalence follows from \eqref{eq:dotHpassivity} and \eqref{eq:tildexi}, while in the third one we have used \eqref{eq:tildeu}. Recalling \eqref{eq:K}, and after some straightforward manipulations, it can be further shown that this is equivalent to:
\begin{equation}
    \dot{\mathcal{V}}_\star=-\begin{bmatrix}
    Q\tilde x\\K_I\tilde  x_c
    \end{bmatrix}^\top \begin{bmatrix}
    \mathcal{R}+\mathcal{K}_P& \frac{1}{2}(g_\star-\bar g)\\
\frac{1}{2}(g_\star-\bar g)^\top &K_L 
    \end{bmatrix} \begin{bmatrix}
    Q\tilde x\\K_I\tilde  x_c
    \end{bmatrix},
\end{equation}
 which, since condition \eqref{eq:condition} hold, is strictly negative. Hence, the equilibrium $(\bar x,\bar  x_c)$ is globally exponentially stable. Finally, if $x_\star\in\mathcal{E}$, $\bar x= x_\star$ and thus $\bar g=\bar g_\star$, from which follows that $\dot{\mathcal{V}}_\star$ is always negative for any $K_P\geq0,K_D\geq 0$ and $K_L>0,$ thus completing the proof.
\end{pf}
 
\begin{rem}\textbf{(Performance limitations of the PLID-PBC).}
In order to establish the performance of the PLID-PBC, consider the case of perfect knowledge of the system parameters, which allows to pick $x_\star\in\mathcal{E}$. Let then $\mathcal Q:=\mathrm{bdiag}(Q^{-1}+\mathcal{K}_D,K_I)\in\mathbb{R}^{n\times n}$ and $\mathcal D:=\mathrm{bdiag}(\mathcal{R}+\mathcal{K}_P,K_L)\in\mathbb{R}^{n\times n}$. Then, there exists an $\alpha_L > 0$ such that 
\begin{equation}
\dot{\mathcal{V}}_\star \leq -\alpha_L \cdot \mathcal{V}_\star(\tilde{x},\tilde{x}_C),\qquad
\alpha_L = 2 \dfrac{\lambda_\mathrm{m}(\mathcal D)}{\lambda_\mathrm{M}(\mathcal Q)},
\end{equation}
\noindent with $\alpha_L$ representing an exponential convergence rate bounding the trajectories of the system. 
\end{rem}

\begin{rem}\label{rem:droop} \textbf{(Droop characteristic of the PLID-PBC).}
Similar to the PID-PBC, the PLID-PBC is unable to exactly regulate the system's states to precise reference values, see also Remark~\ref{rem:deviations}. Nevertheless, deviations from setpoints can be tuned by appropriate design of the leakage---a fact that should be contrasted with the PID-PBC, where these are independent from control parameters. This mechanism is reminiscent of the traditional primary control designs for grid applications, where steady-state values are adjusted by means of \textit{droop}-alike controllers so to achieve an appropriate sharing of the power demand~\cite{vrana,simpson}. It is interesting to note that from the equilibria equations of \eqref{eq:PLIPBC-I}-\eqref{eq:PLIPBC-P}, we have:
\begin{equation}\label{eq:droop}
\bar u-u_\star=D\;(\bar y- y_\star) ,    \end{equation}
with $D:=K_P+K_L^{-1}$, and $\bar y=g(\bar x)^\top Q\bar x= 0_m$, $ y_\star:=  g^\top (x_\star)Q\bar x$. Unsurprisingly, this steady-state relation can be  interpreted as a droop characteristic between the control input $u\in\mathbb{R}^m$ and the passive output $y\in\mathbb{R}^m$, the nature of which is determined by both the proportional gain $K_P$ and the leakage $K_L$.  \end{rem}

\begin{rem} \textbf{(Tuning of the PLID-PBC).}
In contrast with the PID-PBC \eqref{eq:PIPBC-I}-\eqref{eq:PIPBC-P}, using the PLID-PBC \eqref{eq:PLIPBC-I}-\eqref{eq:PLIPBC-P} it is always possible to select the controller gains so to that the corresponding stability conditions are verified. Indeed, by taking $K_P=K_D=0_{m\times m}$ and a sufficiently large leakage $K_L$, \eqref{eq:condition} can be satisfied  independently from the selected reference vector $x_\star\in\hat{\mathcal{E}}$. Hence, the PLID-PBC can be made robust by design to arbitrarily large uncertainties affecting the dissipation matrix $\mathcal{R}$ and the source vector $E$. This is consistent with the fact that as $K_L\rightarrow\infty$, the PLID-PBC closely behaves as a GES open-loop control. We also observe that in presence of perturbations, because of the small values of the dissipation and inertia of the converter, conditions \eqref{eq:K} are verified only for small values of $K_P$ and $K_D$. This fact suggests an intrinsic fragility of the controller to large values of the proportional and derivative gains.
\end{rem}

\begin{rem}\textbf{(GES of generalized PLID controllers).}
We have already seen that the vector $x_\star\in\mathbb{R}^n$, which is provided as a reference to the controller, can be interpreted as an estimation of the equillibrium to be stabilized. However, the proof of Proposition~\ref{prop:PLI} suggests that the matrix $g(x_\star)$ can be replaced in the proportional, integral and derivative channels by any constant matrix $G\in\mathbb{R}^{n\times m}$, and that constant terms can be further introduced, resulting in stability conditions similar to \eqref{eq:condition}. In such case, the controller can be eventually rewritten as:
 \begin{align}
    \dot x_c&=-G^\top Q(x-x_\star)-K_L K_I(x_c-x_{c\star}), \\
    u&=-K_PG^\top Q(x-x_\star)+K_Ix_c-K_DG^\top Q\dot x.
\end{align}
This broader class of controllers includes, among the others, the leaky versions of the PQ controller of Akagi and of the conventional voltage and current PI controllers, which can be all recovered by suitable selection of the matrix $G$.
\end{rem}

\section{Monotone saturating design}\label{sec:saturation}
As anticipated in Remark~\ref{rem:boundU}, the control input physically corresponds to a collection of modulation signals that are constrained to a closed set $\mathcal{U}\subset\mathbb{R}^m$, beyond which they are typically subject to saturation. This may pose a serious problem for the stability of the power converter in closed-loop either with the traditional PID-PBC or with the PLID-PBC. The general question is thus if such controllers are robust to saturation and, if this is not case, whether such robustness can be enforced by properly modifying their design. We try to provide an answer to this question by tackling the problem from a general point of view. Instead of simply considering the saturation as a scalar transformation to be applied to each component of the controller output, we design a suitable $m$-dimensional transformation of the controller that guarantees that its output is maintained at the interior of a given closed set $\mathcal U$. In this frame, the use of conventional scalar saturation function~\cite{rodriguez2000robustly} stands then as a special case. Prior to present such a design, we find convenient to introduce the following definitions. A map $w:\mathbb{R}^m\rightarrow\mathbb{R}^m$ is said to be be strongly monotone if there exists an $\eta>0$ for any $s,h\in\mathbb{R}^m$ such that:
\begin{equation}\label{eq:monotonicity}
    \left[w(s+h)-w(s)\right]^\top h\geq \eta\cdot \vert\vert h\vert\vert ^2.
\end{equation}
 If \eqref{eq:monotonicity} holds  for $\eta=0$ the map is simply said monotone. Note that if $w$ is sufficiently smooth, the following argument further holds~\cite{ryu2016primer}:
 \begin{equation}\label{eq:argumentm}
     \left[w(s+h)-w(s)\right]^\top k=h^\top   \frac{\partial w(s)}{\partial s}k,
 \end{equation}
 for any $k\in\mathbb{R}^m$ and, as a result, \eqref{eq:monotonicity} can be equivalently written as:
  \begin{equation}
     h^\top \frac{\partial w(s)}{\partial s}h\geq \eta\cdot \vert\vert h\vert\vert^2.
 \end{equation}
We now propose to modify the PLID-PBC \eqref{eq:PLIPBC-I}--\eqref{eq:PLIPBC-P}, so that the new controller, in the sequel referred as monotone P leaky I (mPLID)-PBC, is given by:
\begin{align}
    \dot x_c&=-g^\top (x_\star)Qx-K_L \left[ w(K_Ix_c)-w(K_Ix_{c\star})\right],\label{eq:PLIPBC-Im}\\
    v&=-K_Pg^\top (x_\star)Qx+K_Ix_c-K_Dg^\top(x_\star)Q\dot x\label{eq:PLIPBC-Pm}\\
    u&=w(v)\label{eq:PLIPBC-Sm},
\end{align}
where $w:\mathbb{R}^m\rightarrow\mathcal{U}$ is a bounded, sufficiently smooth, strongly monotone map, $x_\star\in\hat{\mathcal{E}}$, $x_{c\star}:=K_I^{-1}\mathsf{u}(x_\star)\in\mathbb{R}^m$, with the map $\mathsf{u}$ given by \eqref{eq:eqcontrol}, are suitably defined reference vectors, and $K_P,K_I,K_D,K_L \in\mathbb{R}^{m\times m}$ are positive semidefinite gain matrices. We have then the following proposition.\smallbreak 

\begin{prop}[GES of the mPLID-PBC]\label{prop:mPLI}
Consider the system \eqref{eq:sys} in closed-loop with the controller
 \eqref{eq:PLIPBC-Im}--\eqref{eq:PLIPBC-Sm}. Assume that there exists an equilibrium point $(\bar x,\bar x_c)\in\mathcal E\times\mathbb{R}^m$ for the closed-loop system, with corresponding equilibrium control $\bar u\in\mathcal{U}$. Define the constant matrices
  \begin{equation}\label{eq:W1W2}
 \begin{aligned}
     M_1:&=\frac{\partial w(s)}{\partial s}\biggr\rvert_{{\bar v}},\qquad M_2:=\frac{\partial w(s)}{\partial s}\biggr\rvert_{{K_I\bar x_c}},
\end{aligned}
 \end{equation}
 where $\bar v:=-K_Pg_\star^\top Q\bar x+K_I\bar x_c$, and let
 \begin{equation}\label{eq:Km}
     \overline{\mathcal{K}}_{P}:=\frac{1}{2}\left[g(\bar x)M_1K_Pg^\top(x_\star)+g(x_\star)K_PM_1g^\top(\bar x)\right],\qquad
    \overline{\mathcal{K}}_{D}:=\frac{1}{2}\left[g(\bar x)M_1K_Dg^\top(x_\star)+g(x_\star)K_DM_1g^\top(\bar x)\right].
 \end{equation}
 Then, if the following inequality hold:
 \begin{equation}\label{eq:conditionm}
    \mathcal{R}+\overline{\mathcal{K}}_P>0,\quad Q^{-1}+\overline{\mathcal{K}}_D>0,\quad M_2K_LM_2>\frac{1}{4}\left[M_2g(x_\star)- M_1g(\bar x)\right]^\top(\mathcal{R}+\overline{\mathcal{K}}_P)^{-1}\left[M_2g(x_\star)- M_1g(\bar x)\right],
\end{equation}
 the equilibrium $(\bar x,\bar x_c)$ is globally exponentially stable and $u(t)\in\mathcal{U}$ for any $t\geq 0$. Moreover, if $x_\star\in\mathcal{E}$ and $w(K_Ix_{c\star})=K_Ix_{c\star}$, then $(x_\star,x_{c\star})\in\mathcal E\times\mathbb{R}^m$ is a globally exponentially stable equilibrium for any positive semidefinite gain matrices $K_P,K_D$ and positive definite gain matrices $K_I,K_L.$ 
 \end{prop}
 \begin{pf}
 Consider the equilibria equations
 \begin{align}
        0_n&=(\mathcal J_0- \mathcal R)Q\bar x+E+g(\bar x)w(\bar v)\label{eq:equilibria1m}\\
        0_m&=-g^\top(x_\star)Q\bar x-K_L\left[w( K_I\bar x_c)-w(K_Ix_{c\star})\right].\label{eq:equilibria2m}
    \end{align}
Then, define $\bar g:=g(\bar x)$, $g_\star:=g(x_\star)$, the incremental variables
\begin{equation}\label{eq:incmon}
    \tilde u:=w(v)-w(\bar v),\quad\tilde v:=v-\bar v,\quad \tilde{x}:=x-\bar x,\quad \tilde{x}_c:=x_c-\bar x_c,
    \end{equation}
    and the incremental energy function  $\mathcal{W}(\tilde x,\tilde x_c):=\mathcal{W}_1(\tilde x)+\mathcal{W}_2(\tilde x_c)$, with
     \begin{equation}
    {\mathcal W}_1(\tilde x):=\mathcal{H}(\tilde x)+\frac{1}{2}\tilde{x}^\top Q\bar g M_1K_Dg_\star^\top Q \tilde {x},\qquad
        \mathcal{W}_2(\tilde x_c):=\int_0^{\tilde x_c}\left[w(K_I\sigma+K_I\bar x_c)-w(K_I\bar x_c)\right]\;\mathrm{d}\sigma.
    \end{equation}
Note that $\tilde{x}^\top Q\bar g M_1K_Dg_\star^\top Q \tilde {x}=\tilde{x}^\top Q\overline{\mathcal{K}}_DQ\tilde x$ and then from \eqref{eq:conditionm} follows that $\mathcal{W}_1$ is positive definite and radially unbounded. Moreover, recalling that $w$ is strongly monotone and $K_I>0$, we have that also $\mathcal{W}_2$ is positive definite and radially unbounded. Hence, we can conclude positive definiteness and radial unboundedness of $\mathcal{W}$. With the definition of incremental variables, from \eqref{eq:PLIPBC-Im}, \eqref{eq:PLIPBC-Sm} and \eqref{eq:equilibria2m} we get:
        \begin{align}
        \dot x_c&=-g_\star^\top Q\tilde x-K_L\left[w(K_Ix_c)-w(K_I\bar x_c)\right]\label{eq:xcm}\\
            \tilde v&=-K_Pg_\star^\top Q\tilde x+K_I\tilde x_c-K_Dg_\star^\top Q\dot x.\label{eq:vtilde}
        \end{align}    
    Hence, by calculating the derivative of $\mathcal{W}$ along the trajectories of the system, we obtain:
    \begin{equation}\label{eq:proofm}
\begin{aligned}
    \dot{{\mathcal{W}}}=&\;\dot{\mathcal{H}}+\tilde{x}^\top Q\bar gM_1K_Dg_\star^\top Q\dot x+\nabla\mathcal{W}_2^\top \dot{\tilde x}_c\\
    =&-\tilde x^\top Q\mathcal{R}Q\tilde x+ \tilde x^\top Q\bar g\left[w(v)-w(\bar v)\right]+\tilde{x}^\top Q\bar gM_1K_Dg_\star^\top Q\dot x+\left[w(K_I x_c)-w(K_I\bar x_c)\right]^\top \dot x_c\\
    =&-\tilde x^\top Q\mathcal{R}Q\tilde x+ \tilde x^\top Q\bar g\left[w(\bar v+\tilde v)-w(\bar v)\right]+\tilde{x}^\top Q\bar gM_1K_Dg_\star^\top Q\dot x-\left[w(K_I \bar x_c+K_I\tilde x_c)-w(K_I\bar x_c)\right]^\top g_\star^\top Q\tilde x+\\
    &-\left[w(K_I \bar x_c+K_I\tilde x_c)-w(K_I\bar x_c)\right]^\top K_L \left[w(K_I \bar x_c+K_I\tilde x_c)-w(K_I\bar x_c)\right]\\
    =&-\tilde x^\top Q\mathcal{R}Q\tilde x+ \tilde x^\top Q\bar g \frac{\partial w}{\partial s}\biggr\rvert_{\bar v}\tilde v+\tilde{x}^\top Q\bar gM_1K_Dg_\star^\top Q\dot x-\tilde x_c^\top K_I \frac{\partial w}{\partial s}\biggr\rvert_{K_I\bar x_c} g_\star^\top Q\tilde x-\tilde x_c^\top K_I \frac{\partial w}{\partial s}\biggr\rvert_{K_I\bar x_c} K_L\frac{\partial w}{\partial s}\biggr\rvert_{K_I\bar x_c} K_I\tilde x_c\\
= &-\tilde x^\top Q\left(\mathcal{R}+\bar gM_1K_Pg_\star^\top\right)Q\tilde x+ \tilde x^\top Q\left(\bar g M_1-g_\star M_2\right)K_I\tilde x_c-\tilde x_c^\top K_IM_2 K_L M_2K_I\tilde x_c,
    \end{aligned}
    \end{equation}
    where in the second equivalence we used \eqref{eq:dotHpassivity} and \eqref{eq:incmon}, in the third we substituted \eqref{eq:xcm}, the fourth follows from argument \eqref{eq:argumentm}, while in the last one we used \eqref{eq:vtilde} and the definitions \eqref{eq:W1W2}. Now recall that by strong monotonicity of $w$, the matrix $M_2$ is positive definite. From \eqref{eq:conditionm} it follows then that $\dot{\mathcal{W}}$ is strictly negative, which implies that the equilibrium $(\bar x,\bar x_c)$ is globally exponentially stable, with boundedness of $u$ trivially following by design of $w$.  Finally, if $x_\star\in\mathcal{E}$, $\bar x= x_\star$ and then $\bar g= g_\star$. At the same time $w(K_Ix_{c\star})=K_Ix_{c\star}$ implies $M_1=M_2$. Hence, $\dot{\mathcal{W}}<0$ for any $K_P\geq 0$ and $K_I>0,K_L>0$, thus completing the proof.
     \end{pf}
\begin{rem}\textbf{(Existence of equilibria under mPLID-PBC).} The existence of an equilibrium $(\bar x,\bar x_c)\in\mathcal{E}\times\mathbb{R}^m$ such that the corresponding equilibrium control $\bar u$ belongs to the set $\mathcal{U}$ is a critical issue for the application of Proposition~\ref{prop:mPLI}. In absence of leakage this is equivalent to the following inclusion:
\begin{equation}
    -g^+(\gamma x_\star)\left[(\mathcal{J}_0-\mathcal{R})Q\gamma x_\star+E\right]\in\mathcal{U},
\end{equation}
which, as we will see in Section~\ref{sec:applications}, might be not verified if $\Delta x=\vert\gamma-1\vert$ takes large values. On the other hand, it can be intuitively understood that increasing the leakage allows a tighter regulation to $u_\star\in\mathcal{U}$ and, as a result, sufficiently large values of $K_L$ eventually guarantee $\bar u\approx u_\star\in\mathcal U.$
\end{rem}
\begin{rem} \textbf{(Robustness of the mPLID-PBC)}. If $w$ is simply monotone, there may exist some equilibrium conditions for which the matrix $M_2$ is only positive semidefinite. As a result, the leakage term might not be able to compensate for the cross-term appearing in the derivative of $\mathcal{W}$. Conversely, if $w$ is strongly monotone, $M_2$ is always positive definite and thus there always exists a sufficiently large $K_L$ that guarantees $\dot{\mathcal{W}}<0.$ In this regard, strong monotonicity can be interpreted as a sufficient condition to preserve the robustness properties already enforced by the leakage.
\end{rem}

 \section{Applications}\label{sec:applications}
\subsection{Dc/dc boost converter}
We consider the problem of voltage regulation of a boost dc/dc power converter interfaced to a  constant impedance, constant current (ZI) load. We proceed by analyzing the implications of the obtained theoretical results, concentrating our attention on the inherent advantages of introducing a leakage in the integral channel. The analysis is validated by simulations on a realistic benchmark.
\subsubsection{Modelling \& preliminary analysis}
\begin{figure}
    \centering
    \includegraphics[width=0.7\columnwidth]{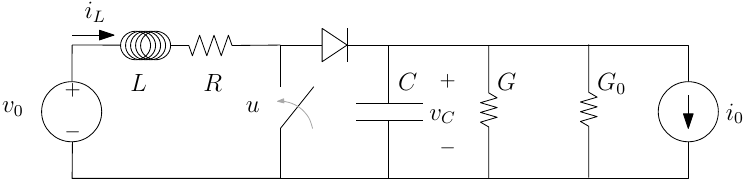}
    \caption{Schematic diagram of a boost converter interfaced to a ZI load.}
    \label{fig:Boost}
\end{figure}

The converter dynamics are captured by the following dynamical system---see Fig. \ref{fig:Boost} for the corresponding electrical scheme:
\begin{equation}\label{eq:dcdc}
\begin{bmatrix}
L\dot i_L\\
C\dot v_C
\end{bmatrix}=
\begin{bmatrix}
-R&-(1-u)\\
1-u&-(G+G_0)
\end{bmatrix}
\begin{bmatrix}
i_L\\
v_C
\end{bmatrix}+
\begin{bmatrix}
v_0\\-i_0
\end{bmatrix},
 \end{equation}
 where: $(i_L,v_C)\in\mathbb{R}^2_{>0}$ are the co-energy variables, respectively the current flowing through the inductor and the voltage across the capacitor; $u\in [u_m\;u_M]\subseteq [0\; 1]$ is the modulation index; $(v_0,i_0)\in\mathbb{R}^2_{>0}$ are the external energy sources, respectively the source voltage and the load current; $R,G$, $L,C$ and $G_0$ are positive constant parameters denoting respectively the resistance, conductance, inductance, capacitance of the converter circuit and the load conductance. 
 By introducing the energy variables $\phi_L:=Li_L$, $q_C:=Cv_C$, it is immediate to obtain the port-Hamiltonian formulation \eqref{eq:sys}, with state vector $x:=\mathrm{col}(\phi_L,q_C)\in\mathbb{R}^2_{>0}$; control input $u\in [u_m\;u_M]\in\mathbb{R}$; source vector $E:=\mathrm{col}(v_0,-i_0)\in\mathbb{R}_{>0}\times\mathbb{R}_{<0}$; interconnection, dissipation and input matrices
\begin{equation}
\mathcal{R}=\begin{bmatrix}
R&0\\0&G+G_0
\end{bmatrix},\quad \mathcal{J}_0=-\mathcal{J}_1=\begin{bmatrix}
0&-1\\1&0
\end{bmatrix},\quad g(x):=\begin{bmatrix}
v_C\\-i_L
\end{bmatrix}
\end{equation}
and Hamiltonian energy function \eqref{eq:hamiltonian} with $Q:=\mathrm{diag}{\{1/L,1/C\}}$. Note that $\mathrm{rank}{(g)}=1$ and thus the system verifies Assumption \ref{ass:rank}. In the sequel, to facilitate physical interpretation, we will refer to the system using co-energy variables. \\
Now assume that the  converter parameters are known and that the voltage source $v_0$ is measurable and sufficiently \textit{stiff}. Using \eqref{eq:powerflow}, the set of assignable equilibria is given by:
\begin{equation}\label{eq:pflowboost} 
    \mathcal{E}:=\lbrace{(i_L,v_C)\in\mathbb{R}^2_{>0}:\;-Ri_L^2-(G+G_0)v_C^2+v_0i_L-i_0 v_C=0\rbrace}, 
\end{equation}
and, for an assignable equilibrium \mbox{$Q\bar x:=\mathrm{col}(\bar i_L,\bar v_C)\in\mathcal{E}$}, the corresponding equilibrium control is:
\begin{equation}\label{eq:boost_baru}
    \bar u=\mathsf{u}(L\bar i_L,C\bar v_C)=1+\frac{R\bar i_L-v_0}{\bar v_C}=1-\frac{(G+G_0)\bar v_C+i_0}{\bar i_L}.
\end{equation}
Since only an estimate of the load parameters is available for the design, the set of \textit{estimated} assignable equilibria does not in general coincide with \eqref{eq:pflowboost}, and is given instead by:
\begin{equation}\label{eq:pflowbooststar}
\hat{\mathcal{E}}:=\lbrace{(i_L,v_C)\in\mathbb{R}^2_{>0}:\;-Ri_L^2-(G+\hat G_0)v_C^2+v_0i_L-\hat i_0 v_C=0\rbrace}, 
\end{equation}
where $\hat G_0\in\mathbb{R}_{>0}$, and $\hat i_0\in\mathbb{R}_{>0}$ denote the estimated value of the load conductance and current respectively. Accordingly, the estimated passive output \eqref{eq:passiveoutput} reads:
\begin{equation}
 y=v_{C\star} i_L-i_{L\star} v_C,
\end{equation}
 where $x_\star:=(i_{L\star},v_{C\star})\in\hat{\mathcal{E}}$.\\
 
 \subsubsection{Control design}
The control objective is to guarantee that the output voltage $v_C$ is regulated to its nominal value $v_{C\star}$ in nominal operating conditions. Moreover, the controller must be equipped with an additional  control mechanism that allows to adjust the voltage deviations by appropriate tuning of a gain, in case of perturbed conditions stemming by an inaccurate knowledge of the load current $i_0$ and conductance $G_0$. The definition of an appropriate rationale for the selection of such gain may vary depending on the considered application and is left for future investigation. Before proceeding any further, it is important to note from \eqref{eq:boost_baru} that, because of the typically small values of $R$, we have:
\begin{equation}
\bar v_C\approx \frac{v_0}{1-\bar u}.
\end{equation}
Hence, under this approximation the voltage is a simple function of the modulation index at steady-state. Consequently, tight regulation of $u$ to $u_\star^{\mathrm{approx}}:=1-v_0/v_{C\star}$ ensures approximate regulation of $v_C$ to $v_{C\star}$. Whenever voltage deviations are thus required to be very small, one might be tempted to simply apply a constant control $u_\star^{\mathrm{approx}}$ to fulfill such objective. However, it must be recalled that this would trigger large oscillations in the controlled system---see Remark~\ref{rem:minper}---a fact that motivates the use of a feedback action.\smallbreak

\begin{figure}
    \centering
    \includegraphics[width=0.45\columnwidth]{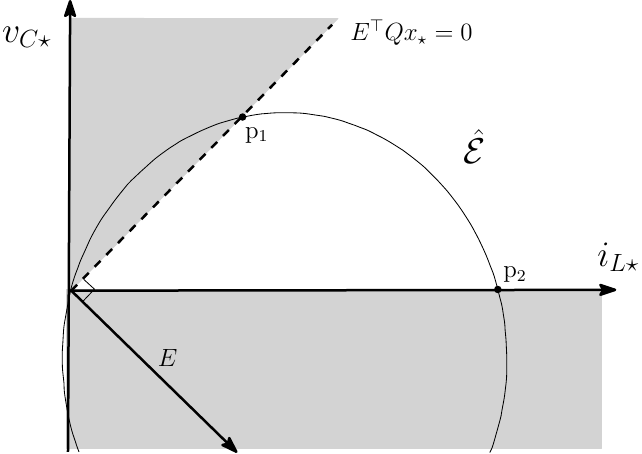}  
    \caption{Graphical interpretation of the stability condition $P_\mathrm{net}(x_\star)>0$ for the boost converter.
    }
    \label{fig:Boost_graph}
\end{figure}

\noindent\textit{PID-PBC.}  As per Proposition~\ref{prop:gamma},  the robustness margins of the PID-PBC are established via the condition $P_\mathrm{net}(x_\star)=E^\top Qx_\star>0$, which is independent from the controller parameters.  An interesting geometric interpretation of this condition is that it corresponds to have the point $x_\star\in\hat{\mathcal{E}}$ contained in the half-plane orthogonal to $E$.  The set of estimated  equilibria that verify the stability condition are thus given by the arc of the circle $\hat{\mathcal{E}}$ connecting clockwise the point $\mathrm{p}_1$ to the point $\mathrm{p}_2$---see also Fig.~\ref{fig:Boost_graph}.
 According to Proposition~\ref{prop:gamma}, and using the fact that $x_\star\in\hat{\mathcal{E}}$ the following analytical, stability condition can be then established:
  \begin{equation}\label{eq:stabilityboost}
     (i_0-\hat i_0)v_{C\star}+(G_0-\hat{G}_0)v^2_{C\star}<Ri_{L\star}^2+(G+ G_0)v_{C\star}^2,
 \end{equation}
together with a characterization of the steady-state deviations via the following scalar---see also Remark~\ref{rem:deviations}:
 \begin{equation}\label{eq:deltaboost}
     \Delta x=\vert\gamma-1\vert=\Bigg\vert\dfrac{v_0i_{L\star}-v_{C\star}i_0}{Ri_{L\star}^2+(G+G_0)v_{C\star}^2}-1\Bigg\vert\leq \frac{\vert\hat i_0-i_0\vert v_{C\star}+\vert\hat G_0-G_0\vert v_{C\star}^2}{Ri_{L\star}^2+(G+G_0)v_{C\star}^2} .
 \end{equation}
It is then immediate to see from \eqref{eq:stabilityboost}-\eqref{eq:deltaboost} that, as the load current and load conductance estimation error increase, deviations from the desired operating point $x_\star$ are exacerbated and the stability condition is eventually violated. Note that such situation is more likely to occur if the system is lightly damped, as for the case of pure current loads. All in all we can draw the conclusion that both robustness margins and steady-state deviations are strongly affected by the estimation errors and the inherent ability of the system to absorb them by means of its dissipative components.  \smallbreak

\noindent\textit{PLID-PBC.} With the introduction of a leakage in the integral channel, robustness can be improved and deviations limited even for lightly damped systems, provided that an appropriate tuning of the control parameters is realized. For illustrative purposes, let us analyze the case where $K_P=K_D=0$, so to provide an insightful interpretation of the robust stabilizing action of the leaky integral channel only. In such case, from Proposition~\ref{prop:PLI} we have that the stability of the equilibrium is ensured simply if:
\begin{equation}\label{eq:conditionboost}
    K_L>\frac{1}{4RG}\left[R(i_{L\star}-\bar i_L)^2+G(v_{C\star}-\bar v_C)^2\right],
\end{equation}
being $\mathcal{R}+\mathcal{K}_P=\mathcal{R}>0.$ As a result,  it is sufficient to design the leakage large enough to compensate the weighted quadratic error between the actual and the estimated equilibrium point. While it can be verified that large values of $K_P$ and $K_D$ may eventually not satisfy the stability conditions \eqref{eq:K}, it is important to recall that the proportional and derivative gains strongly affect the performances of the closed-loop system and thus, as for the leakage $K_L$, this must be carefully tuned in order to obtain a good compromise between robustness and performance.\smallbreak

\noindent\textit{mPLID-PBC.} The control input established via the PID- and PLID-PBC may eventually take large values, a fact that is clearly not acceptable in practice, since the modulation index is constrained to the set $\mathcal{U}:=[u_m\;u_M]\subseteq[0\;1]$. To guarantee that the control input is maintained within such bounds we complement the design by formulating a mPID- and mPLID-PBC as in \eqref{eq:PLIPBC-Im}--\eqref{eq:PLIPBC-Sm}, with strongly monotone function $w:\mathbb{R}\rightarrow[u_m\;u_M]$ given by:
 \begin{equation}\label{eq:wboost}
     w(s):=\frac{u_M-u_m}{2}\cdot \tanh (\lambda s-u_0)+\frac{u_M+u_m}{2},\qquad u_0:=\lambda u_\star+\tanh^{-1}\left(\frac{u_M+u_m-2u_\star}{u_M-u_m}\right),
 \end{equation}
 where $\lambda>0$ is a design parameter denoting the steepness of such a function. This design guarantees by definition that the modulation index is maintained within the prescribed bounds and that $w(u_\star)=u_\star$, that is, the monotone map preserves the nominal reference $u_\star$.  
\subsubsection{Simulations} 
\begin{figure}[ht]
    \centering
    \includegraphics[width=0.49\columnwidth]{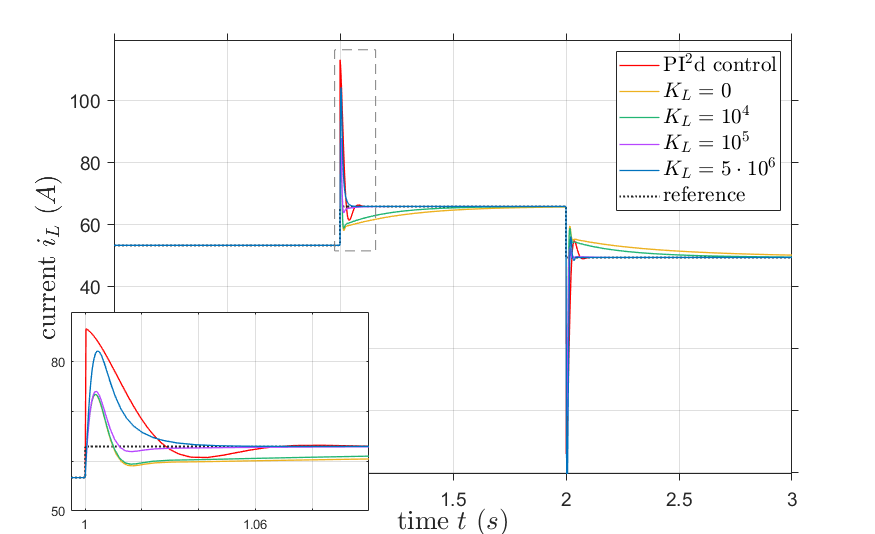}
     \includegraphics[width=0.49\columnwidth]{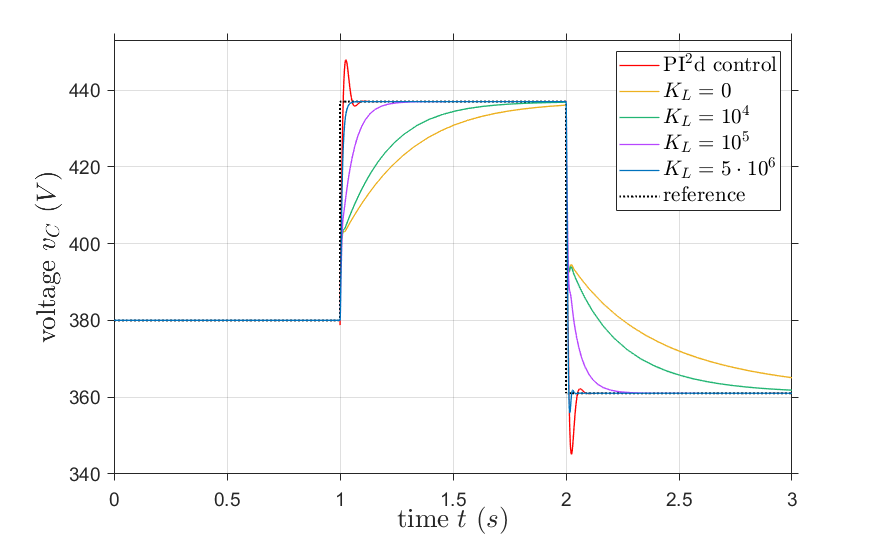}
    \caption{Current and voltage responses of the boost converter under mPLID-PBC \eqref{eq:PLIPBC-Im}-\eqref{eq:PLIPBC-Sm}, with $w$ given by \eqref{eq:wboost} and PI$^2$d, in nominal conditions. For the tuning of the mPLID-PBC we set $\lambda=1$, $K_P=10^{-5}$, $K_I=10^{-3}$, $K_D=10^{-9}$ and progressively increased the value of the leakage.}
    \label{fig:Boost_nominal}
\end{figure}

\begin{figure}[ht]
   \centering
   \includegraphics[width=0.49\columnwidth]{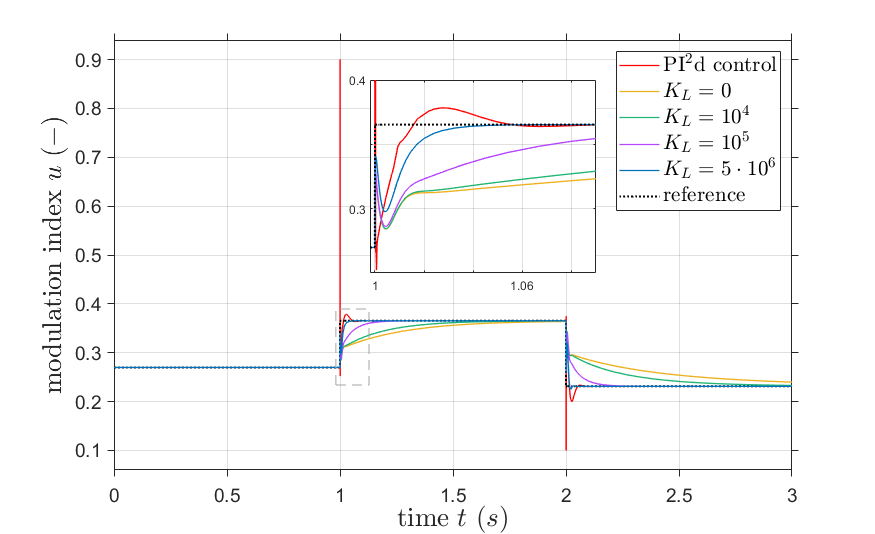}
   \caption{Modulation index responses of the boost converter under mPLID-PBC \eqref{eq:PLIPBC-Im}-\eqref{eq:PLIPBC-Sm}, with $w$ given by \eqref{eq:wboost} and PI$^2$d, in nominal conditions. For the tuning of the mPLID-PBC we set $\lambda=1$, $K_P=10^{-5}$, $K_I=10^{-3}$, $K_D=10^{-9}$ and progressively increased the value of the leakage.}
    \label{fig:Boost_nominal_U}
\end{figure}

\begin{figure} 
    \includegraphics[width=0.49\columnwidth]{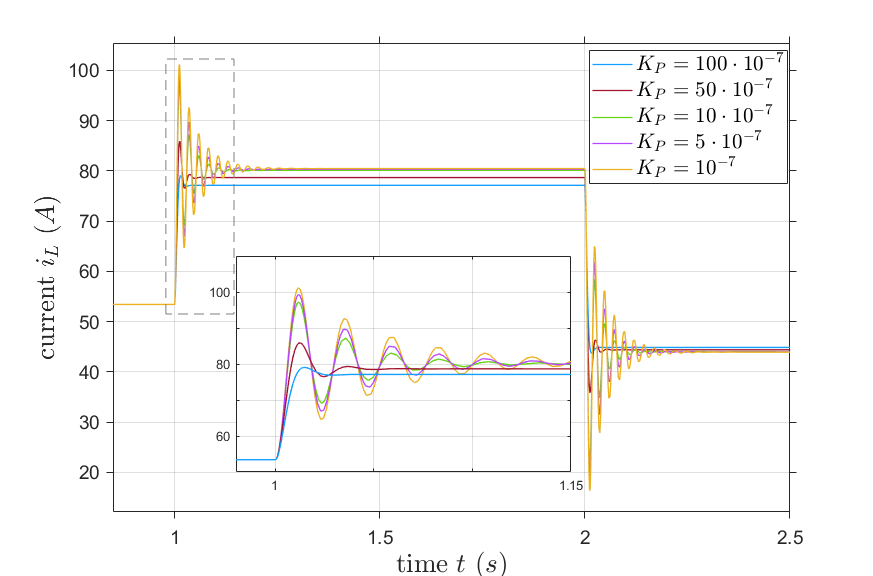}
    \includegraphics[width=0.49\columnwidth]{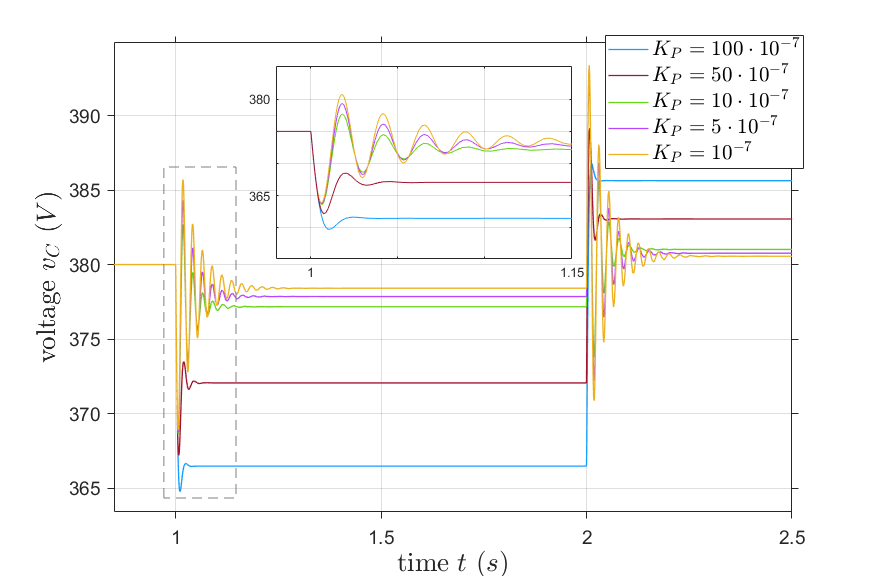}
    \caption{Current and voltage responses of the boost converter under mPLID-PBC \eqref{eq:PLIPBC-Im}-\eqref{eq:PLIPBC-Sm}, with $w$ given by \eqref{eq:wboost}, in perturbed conditions. For the tuning we set a $\lambda=1$, $K_L=5\cdot 10^{6}$, $K_I=10^{-3}$, $K_D=10^{-9}$ and progressively decreased the value of the proportional gain.}
    \label{fig:Boost_Perturbed}
\end{figure}

\begin{figure}[ht]
\centering
 \includegraphics[width=0.49\columnwidth]{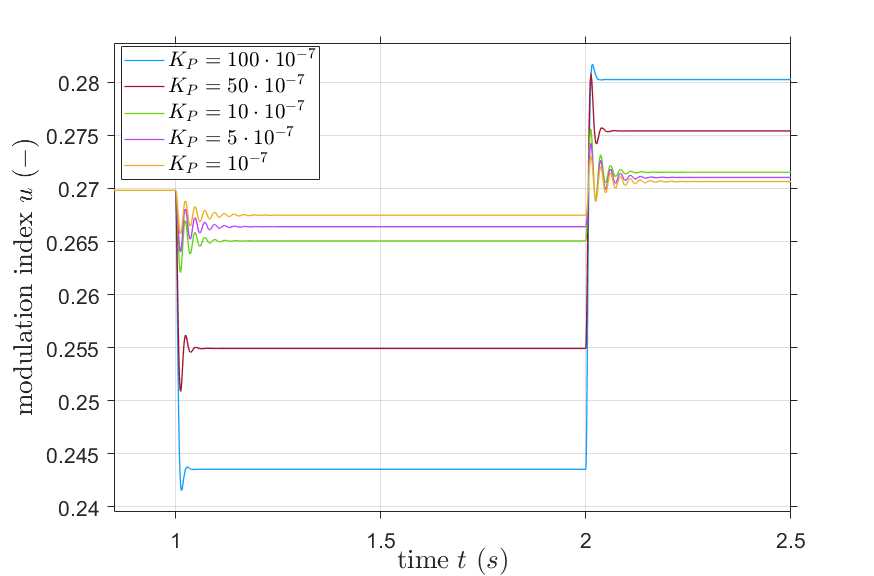}
 \includegraphics[width=0.49\columnwidth]{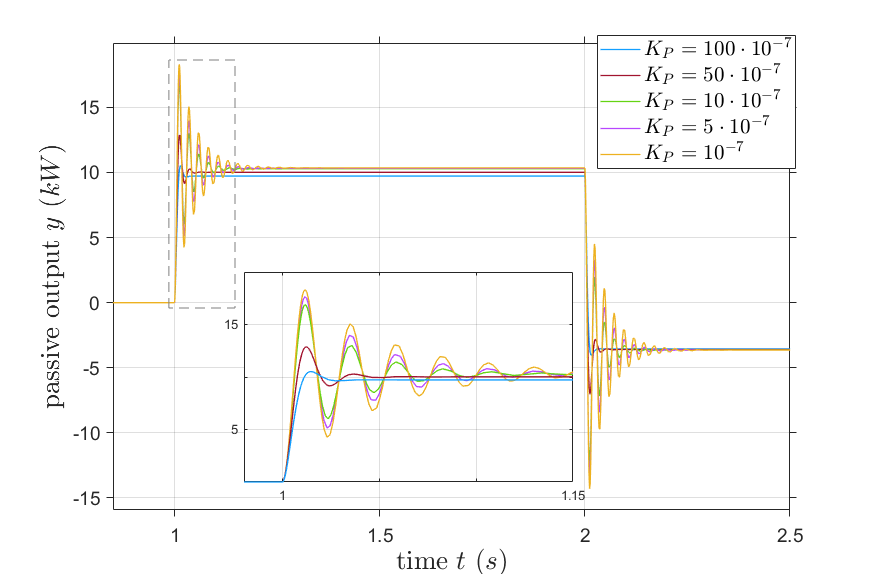}
 \caption{Modulation index and passive output responses of the boost converter under mPLID-PBC \eqref{eq:PLIPBC-Im}-\eqref{eq:PLIPBC-Sm}, with $w$ given by \eqref{eq:wboost}, in perturbed conditions. For the tuning we set $\lambda=1$, $K_L=5\cdot 10^{6}$, $K_I=10^{-3}$, $K_D=10^{-9}$ and progressively decreased the value of the proportional gain.}
    \label{fig:Boost_Perturbed_U}
\end{figure}

\begin{figure}[ht]
   \centering
   \includegraphics[width=0.49\columnwidth]{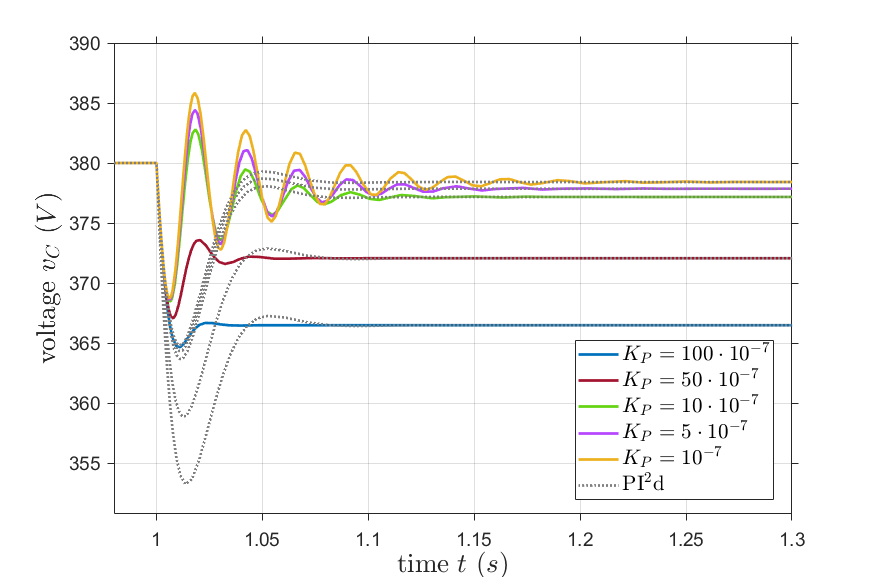}
   \caption{Detail of the voltage responses of the boost converter under mPLID-PBC \eqref{eq:PLIPBC-Im}-\eqref{eq:PLIPBC-Sm}, with $w$ given by \eqref{eq:wboost} and compared with a PI$^2$d, in perturbed conditions. For the tuning of the mPLID-PBC we set $\lambda=1$, $K_L=5\cdot 10^{6}$, $K_I=10^{-3}$, $K_D=10^{-9}$ and progressively decreased the value of the proportional gain. }
    \label{fig:Boost_droops}
\end{figure}

To validate our considerations via simulations we assume that the boost converter is lightly damped and characterized by the parameters given in Table~\ref{table:boost_parameters}. We consider two scenarios and evaluate the responses under mPID-, mPLID-PBC and a traditional cascaded control scheme, including an inner current PI controller and an outer voltage PI controller further equipped with a linear droop mechanism---in the sequel referred as PI$^2$d. For a first wave of simulation we assume that the load current and conductance are known, with values $\hat G_0=40\;mS$, $\hat i_0=20\;A$, and that the desired output voltage $v_{C}^\star$ is modified by $15\%$ and $5\%$ of its initial, nominal value of $v_{C}^\star(0)=380\; V$. Since these variations are imposed by the user, it is possible to simultaneously compute, using \eqref{eq:pflowboost} and \eqref{eq:boost_baru}, the corresponding 
references $x_\star=\mathrm{col}(i_{L\star},v_{C\star})\in\mathcal{E}$, $x_{c\star}=u_\star/K_I$  to be provided to the controller.
\begin{table}[t]
    \caption{Boost converter and nominal load parameters \cite{cucuzzella}.}\label{table:boost_parameters}
    \centering
    \begin{tabularx}{\textwidth}{cc|cc|cc|cc|cc|cc|cc|cc}
    \toprule
     $L$ & $1.12\;mH$ & $R$ & $10\;m\Omega$ &    $C$ & $6.8\phantom{0}\;mF$ & $G$ &  {$10\;mS$}  & $v_0$ & $278\;V$ & $\hat i_0$ & $\phantom{0}20\;A$  &  $u_m$ & $0.1$ &  $u_M$ & $0.9$
     \end{tabularx} 
\end{table}
The responses are illustrated in Figs. \ref{fig:Boost_nominal}-\ref{fig:Boost_nominal_U}, assuming that voltage reference changes at $1\;s$ and $2\;s$. It can be seen that  the the mPID-PBC preserves stability under saturation, but exhibits poor performances, that cannot be further improved by appropriate tuning of the proportional, integral and derivative gains, which are set at $K_P=10^{-5}$, $K_I=10^{-3}$ and $K_D=10^{-9}$ respectively. On the other hand, the introduction of a leakage speeds up convergence and further avoids saturation---in accordance with the obtained theoretical results---obtaining even better performances than the standard PI$^2$d controller.
For the second wave of simulations, we assume that the converter is regulating its voltage to the nominal value $v_C^\mathrm{nom}$ and consider inadvertent changes in the load current at $T=1\;s$ and  $2T$. More precisely we assume that at time $T$ a $+100\%$ change occurs in the load current, followed by a drop of $-65\%$ at time $2T$. In order to assess practical implementability of the controller in absence of leakage, the corresponding values of $\bar u$ and  $\Delta x$ are evaluated, leading to the following considerations. First, the steady-state output generated by the controller is not always attainable, since $\bar u\notin\mathcal U$ from $T$ to $2T$, systematically triggering saturation of its output. Second, from $2T$ to $3T$ large deviations are expected, as $\Delta x$ takes a large value. Since in such situations the responses are observed to evolve far away from the values of physical interest, we decide to not report them in this paper. Instead, the responses of the system under mPLID-PBC are illustrated in Figs.~\ref{fig:Boost_Perturbed}-\ref{fig:Boost_Perturbed_U}.
For such design, we select a sufficiently large value of the leakage $K_L=5\cdot 10^{6}$, which in nominal conditions was shown to ensure a quick settling time. Moreover, responses are evaluated for different values of the proportional gain $K_P$ that verify the corresponding stability condition. It can be seen that in all simulated scenarios the trajectories quickly converge to steady-state, with current closely following the load demand. Unsurprisingly, a tighter regulation of the voltage is achieved for smaller values of the proportional gains, which result in a nearly horizontal droop slope, \textit{i.e.} $D=K_P+1/K_L\approx 0$, see also Remark~\ref{rem:droop}. However, when the value of the proportional gain is too small, this may eventually trigger large oscillations, which can be explained by the fact that in such condition the controller behaves similarly to an open-loop control. We conclude then that an accurate tuning must be realized in order to obtain a good compromise between performances and steady-state deviations. To better contextualize our results, in Fig.~\ref{fig:Boost_droops} we further compare the voltage responses under mPLID-PBC and PI$^2$d. It is shown that appropriate selection of the leakage allows to recover specific steady-state droop characteristics imposed by the controller PI$^2$d, and that performances of the mPLID-PBC eventually outperform the performances of the  PI$^2$d as the value of the droop is increased. This fact, together with the availability of closed form stability conditions, suggests that the mPLID-PBC should be considered as a competitive alternative to more traditional primary control solutions. 

 \subsection{HVDC grid-connected Voltage Source Converter}
 We consider a two-level voltage source converter (2L-VSC) in single-terminal HVDC configuration that requires adequate regulation of the active and reactive power to desirable values. Due to limited space available, for such application we focus exclusively on the design of a mPID-PBC, \textit{i.e.} with no leakage, thus proving that such a design may remain of some interest for specific modes of operations of the power converters. As for the case of the boost converter, the analysis of the theoretical results is complemented  by detailed simulations.
 \subsubsection{Modelling \& preliminary analysis}
 For the modelling of the system we assume that the ac side of the 2L-VSC is connected to a \textit{stiff} grid, with fixed voltage amplitude and frequency, the values of which are readily available for the control design. This can be justified by the usually fast operation of the synchronization mechanisms, such as phase-locked loops, compared to the rate of variation of the frequency, which further allows for a representation of the system in a suitable $dq$-frame~\cite{erickson2007fundamentals}. To provide a more realistic description of the power converter in a grid setting, we also assume that it corresponds to one of the terminals ($\mathsf{t1}$) of a point-to-point HVDC transmission system, whose model is completed by a single transmission line interconnecting a dc voltage source. This is a common modelling practice, where the dc voltage source is representative of a second terminal ($\mathsf{t2}$) constituted by another 2L-VSC operated in dc voltage-controlled mode~\cite{PEMC2014}.
  \begin{figure}[ht]
    \centering
     \includegraphics[width=0.8\columnwidth]{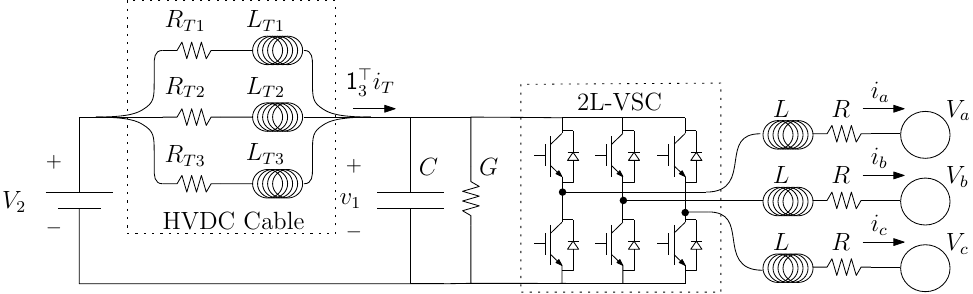}
    \caption{Schematic diagram of the point-to-point HVDC transmission system under study.}
    \label{fig:2LVSC}
\end{figure}
 The circuit diagram of the interconnected system is illustrated in Fig. \ref{fig:2LVSC}. The terminal $\mathsf{t1}$ is then described by the following dynamical system:
\begin{equation}
    \begin{bmatrix}
    L\dot{i}_{d}\\
    L\dot{i}_{q}\\
    C\dot{v}_1
    \end{bmatrix} =
    \begin{bmatrix}
    -R &-L\omega &u_{d}\\
    L\omega  & -R &u_q \\
    -u_{d}&-u_{q}&-G
        \end{bmatrix}
    \begin{bmatrix}
    i_{d}\\
    i_{q}\\
    v_1\\
    \end{bmatrix} + 
    \begin{bmatrix}
    {-}V_{d}\\
    0\\
    i_{0}
        \end{bmatrix}
\end{equation}
\noindent where $i_{dq}:=\mathrm{col}(i_{d},i_{q})\in\mathbb{R}^2$,  $v_{1}\in\mathbb{R}_{>0}$ are the co-energy variables, respectively the direct and quadrature currents flowing through the inductor and the dc  voltage across the capacitor; $u_{dq}:=\mathrm{col}(u_d,u_q)\in\mathcal U\subset\mathbb{R}^m$ are the active and reactive components of the modulation indices;  $v_{0}:=\mathrm{col}(V_{d},V_{q})\in \mathbb{R}_{>0}^2$, with $V_q=0$, are the external ac energy sources, respectively the direct and quadrature components of the ac grid voltage; $i_{0}\in\mathbb{R}$ is the dc current entering terminal $\mathsf{t1}$; $R$, $G$, $L$, $C$ and $\omega$ are positive constant parameters denoting respectively the resistance, conductance, inductance and capacitance of the converter circuit and the frequency of the ac grid. Following standards in the modelling of HVDC cables, we assume that the transmission line is described by three parallel $RL$ branches, whose parameters are established via an appropriate vector-fitting procedure~\cite{CableI,CableII}. The line dynamics are captured by the following linear differential equation:
\begin{equation}
    L_T\dot i_T=-R_Ti_T+\mathsf{1}_3(V_2-v_1),
\end{equation}
    \noindent where: $i_T\in\mathbb{R}^3$ is the vector of currents flowing through the $RL$ branches; $V_2\in\mathbb{R}_{>0}$ is the constant dc voltage source imposed at terminal $\mathsf{t2}$; $R_T\in\mathbb{R}^{3\times 3}$, $L_T\in\mathbb{R}^{3\times 3}$ are positive diagonal matrices including the line resistances and inductances. By introducing the energy variables $\phi_d:=Li_d$, $\phi_q:=Li_q$, $q_1:=Cv_1$, $\phi_T:=L_Ti_T$ and recalling from Fig.~\ref{fig:2LVSC} that the dc current is such that $i_0=\mathsf{1}_3^\top i_T$, it is immediate to obtain the port-Hamiltonian formulation~\eqref{eq:sys} with state vector $x:=\mathrm{col}(\phi_d,\phi_q,q_1,\phi_T)\in\mathbb{R}^2\times\mathbb{R}_{>0}\times\mathbb{R}^3$; control input $u\in\mathcal U$; source vector $E:=\mathrm{col}(V_d,0,0,\mathsf{1}_3V_2)\in\mathbb{R}_{\geq 0}^6$; interconnection, dissipation matrices:
\begin{equation}
    \mathcal{J}_0=L\omega J_{21}+J_{43}+J_{53}+J_{63} , \quad
    \mathcal{J}_1=J_{13}, \quad
    \mathcal{J}_2=J_{23}, \quad
    \mathcal{R}:=\mathrm{bdiag}\{R\mathbb{I}_2,G,R_T\},
\end{equation}
where $J_{ik}\in\mathbb{R}^{6\times 6}$ is a skew-symmetric matrix with entries $(i,k)=1$, $(k,i)=-1$ and zero elsewehere, and input matrix
\begin{equation}
g(x):=
    \begin{bmatrix}
    v_1\mathbb{I}_2\\
    -i_{dq}^\top\\
    0_{3\times 2}
    \end{bmatrix}\in\mathbb{R}^{6\times 2}.
\end{equation}    
\noindent The Hamiltonian energy function is then given by \eqref{eq:hamiltonian} with  $Q:=\mathrm{bdiag}\lbrace\frac{1}{L}\mathbb{I}_2,\frac{1}{C},L^{-1}_T\rbrace$. In view of the $dq$-frame transformation, the control input vector is constrained to the square $\mathcal{U}:=\left[-\frac{2}{3}\;\frac{2}{3}\right]^2\subset\mathbb{R}^2$. We further assume that converter and line parameters are known and that the ac voltage source is measurable. The dc voltage-controlled terminal $\mathsf{t2}$ is subject to constant perturbations, which may arise following a variation of the power demand from the corresponding ac side.  As for the boost converter, we prefer to adopt a description in co-energy variables. Using \eqref{eq:powerflow}, the set of assignable equilibria is given then by:
\begin{equation}\label{eq:pflowvsc} 
\begin{aligned}
    \mathcal{E}:=\{(i_d,i_q,v_1,i_T)\in\mathbb{R}^2\times\mathbb{R}_{>0}\times\mathbb{R}:\; i_T&=G_T\mathsf{1}_3( V_2-v_1)\;\mathrm{and}\\0&=-R(i_d^2+i_q^2)-(G+\mathsf{1}_3^\top G_T\mathsf{1}_3)v_1^2{-}V_di_d+\mathsf{1}_3^\top G_T\mathsf{1}_3v_1V_2\}, 
    \end{aligned}
\end{equation}
 with $G_T:=R^{-1}_T$, while, for an assignable equilibrium \mbox{$Q\bar x:=\mathrm{col}(\bar i_d,\bar i_q,\bar v_1,\bar i_T)\in\mathcal{E}$}, the corresponding equilibrium control is:
\begin{equation}\label{eq:vsc_baru}
    \bar u_d=\frac{1}{\bar v_1}\left(R\bar i_d+L\omega \bar i_q{+}V_d\right).\qquad \bar u_q=\frac{1}{\bar v_1}\left(R\bar i_q-L\omega \bar i_d\right)
\end{equation}
However, since the voltage $V_2$ is subject to perturbations, only a set of \textit{estimated} assignable equilibria is available for the design, and is given by:
\begin{equation}\label{eq:pflowvscstar}
\begin{aligned}
    \hat{\mathcal{E}}:=\{(i_d,i_q,v_1,i_T)\in\mathbb{R}^2\times\mathbb{R}_{>0\times\mathbb{R}}:\; i_T&= G_{T}\mathsf{1}_3( \hat V_2- v_1)\;\mathrm{and}\\0&=-R(i_d^2+i_q^2)-(G+\mathsf{1}_3^\top G_{T}\mathsf{1}_3)v_1^2{-}V_di_d+\mathsf{1}_3^\top G_{T}\mathsf{1}_3 v_1\hat V_2\}, 
    \end{aligned}
\end{equation}
where $\hat V_2\in\mathbb{R}_{>0}$ denotes the estimated value of the dc voltage source. Accordingly, the estimated passive output \eqref{eq:passiveoutput} reads:
\begin{equation}
 y_d=v_{1\star} i_d-i_{d\star} v_1,\quad y_q=v_{1\star} i_q-i_{q\star} v_1,
\end{equation}
 where $(i_{d\star},i_{q\star},v_{1\star},i_{T\star})\in \hat{\mathcal{E}}$. \\
 
\subsubsection{Control design}
We restrict our attention to the regulation of active and reactive power that, in reason of the $dq$ frame adopted, are given by:
$$
P:={\dfrac{3}{2}}V_di_d,\qquad Q:={\dfrac{3}{2}}V_di_q.
$$
Hence, since $V_d$ is constant, their regulation is equivalent to the regulation of the direct current $i_d$ and quadrature current $i_q$ respectively. The quadrature current needs to be regulated near to $i_{q\star}$ and the direct current must converge close to a nominal value $i_{d\star}$. Such objectives must be fulfilled even in case of fluctuations of the voltage $V_2$, which is independently controlled at terminal $\mathsf{t2}$. \smallbreak

 \noindent\textit{PID-PBC.}  Although the the system does not verify Assumption \ref{ass:rank}, in view of the linearity of the transmission line circuit a result similar to the one obtained in Proposition~\ref{prop:gamma} can be derived. Indeed, with analogous calculations it is possible to show that:
 $$
 \begin{bmatrix}
 \bar i_{dq}\\\bar v_1
 \end{bmatrix}=\gamma
 \begin{bmatrix}
 i^\star_{dq}\\ v^\star_1
 \end{bmatrix},\qquad \gamma:=\frac{{-}V_di_{d\star}+\mathsf{1}_3^\top G_T\mathsf{1}_3v_{1\star}V_2}{R(i_{d\star}^2+i_{q\star}^2)+(G+\mathsf{1}_3^\top G_{T}\mathsf{1}_3)v_{1\star}^2}.
 $$
and that stability is guaranteed if $\gamma>0$. Since the denominator is always positive, and recalling \eqref{eq:pflowvscstar}, this is equivalent to:
 \begin{equation}
 \hat V_2- V_2<\frac{R(i_{d\star}^2+i_{q\star}^2)+(G+\mathsf{1}_3^\top G_T\mathsf{1}_3)v_{1\star}^2}{\mathsf{1}_3^\top G_T\mathsf{1}_3v_{1\star}}.
 \end{equation}
Moreover, steady-state deviations from the desired equilibrium are captured by the value of $\Delta x$ that, using again \eqref{eq:pflowvscstar}, can be rewritten as:
 \begin{equation}\label{eq:deltaX_vsc}
 \Delta x=\vert\gamma-1\vert=
 \frac{ \mathsf{1}_3^\top G_T\mathsf{1}_3v_{1\star} }{R(i_{d\star}^2+i_{q\star}^2)+(G+\mathsf{1}_3^\top G_{T}\mathsf{1}_3)v_{1\star}^2}\cdot \vert V_2-\hat V_2\vert,
\end{equation}
from which immediately follows that both stability margins and deviations from the desired equilibria are in strict relation with the error between the actual and nominal dc voltage source. Therefore, we can conclude that the practical implementability of the PID-PBC is strictly linked to the ability of the grid to guarantee a tight voltage control at terminal $\mathsf{t2}$.  It is also interesting to note that deviations are mitigated or exacerbated by the \textit{stiffness} of the transmission system. More precisely, large (respectively small) deviations are expected for small (respectively large) values of $G_T$. These considerations are consistent with the practical operation of point-to-point HVDC transmission systems, where it is common to define a \textit{slave} terminal, which sets the active and reactive power demand/supply, and a \textit{master} terminal, which regulates the dc voltage to ensure that the resulting power balance is satisfied~\cite{vrana}.\\
At this point it is convenient to make some practical considerations in relation to the control objectives and the real parameters of an HVDC transmission system. First, note that the traditional objective of regulating the reactive power $Q$ to zero implies $\bar i_q=\gamma i_{q\star}=0$, regardless of the value of $\gamma$. Hence, zero reactive power regulation can be guaranteed using a PID-PBC independently from perturbations. Second, converter losses are negligible compared to the losses of the HVDC line.  As a result, the following approximate relation holds:
\begin{equation}
 \Delta x\approx \frac{\vert V_2-\hat V_2\vert}{v_{1\star}}.
 \end{equation}
Since the voltage perturbations are typically small compared to the high voltage operation value of HVDC systems, we expect $\Delta x$ to be sufficiently small---a fact that will be illustrated in simulations.\smallbreak

\noindent\textit{mPID-PBC.} To cope with the problem of possible saturation of the control input we consider the monotone modification of the controller discussed in Section~\ref{sec:saturation}.
The mPID-PBC is thus designed according to \eqref{eq:PLIPBC-Im}--\eqref{eq:PLIPBC-Sm}, with strongly monotone map $w_{dq}:\mathbb{R}^2\rightarrow\mathcal{U}$ given by $w_{dq}:=\mathrm{col}(w(s_d),w(s_q))$ where $w$ is the same monotone function employed for the boost converter with $u_M=u_m=2/3$, see~\eqref{eq:wboost}.

\subsubsection{Simulations} 
\begin{figure}
    \centering
    \includegraphics[width=0.49\columnwidth]{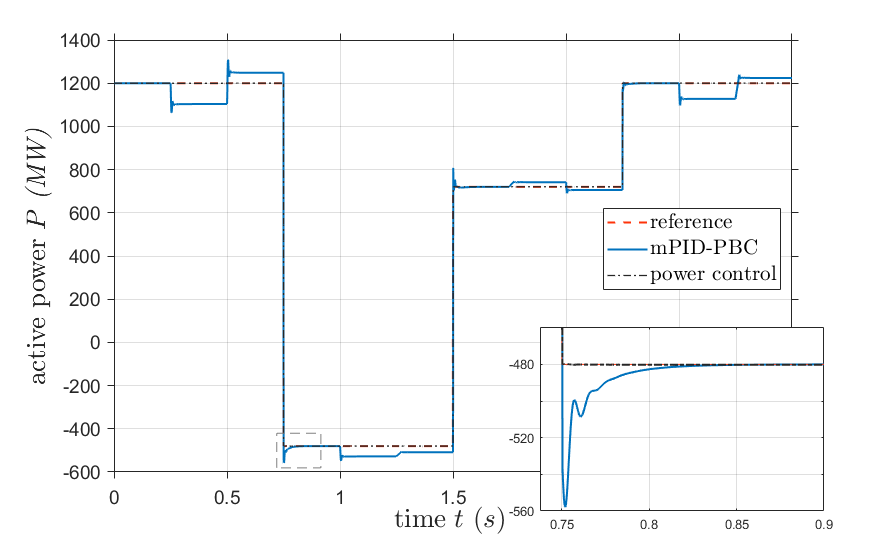}
    \includegraphics[width=0.49\columnwidth]{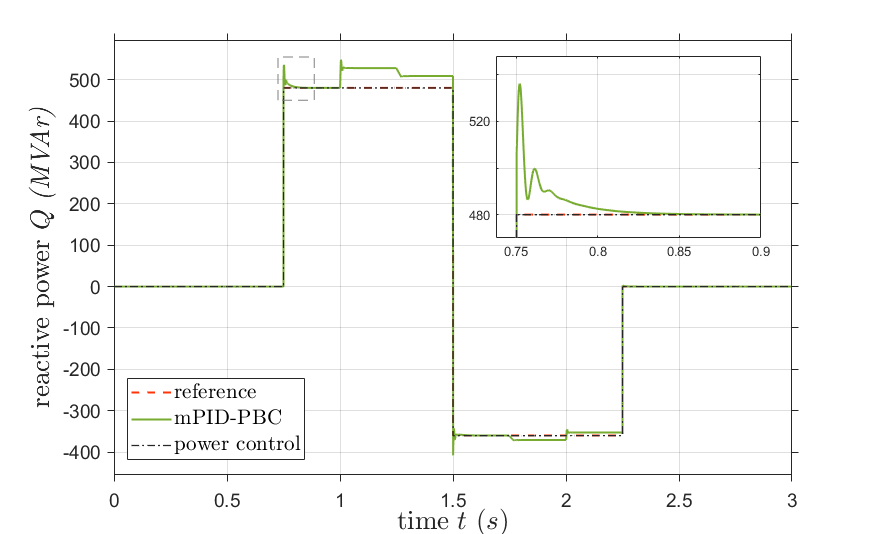}
    \caption{Active and reactive power responses of the 2L-VSC under mPID-PBC \eqref{eq:PIPBC-I}-\eqref{eq:PIPBC-P}, with $w$ given by \eqref{eq:wboost}, for the considered scenario.}
    \label{fig:2LVSC_idq}
\end{figure}

\begin{figure}
    \centering
    \includegraphics[width=0.49\columnwidth]{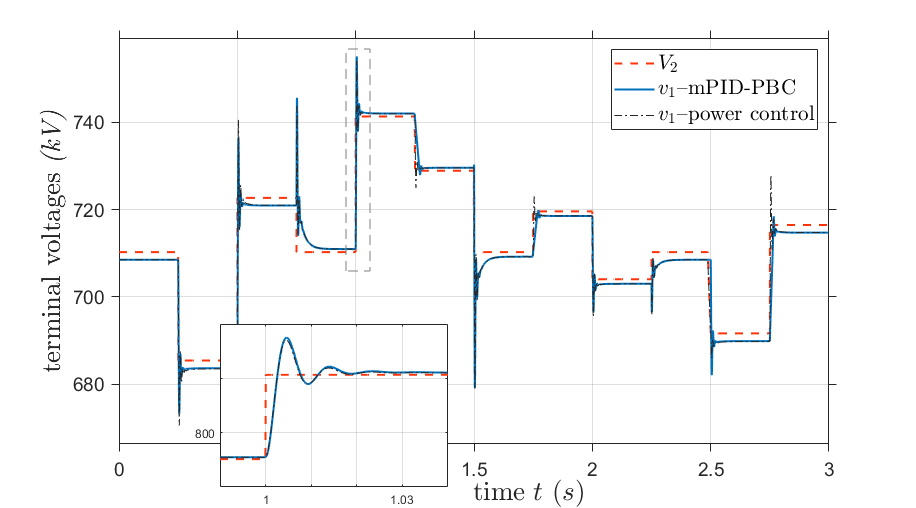}
    \caption{Terminal voltages under mPID-PBC \eqref{eq:PIPBC-I}-\eqref{eq:PIPBC-P}, with $w$ given by \eqref{eq:wboost}, for the considered scenario.}
    \label{fig:2LVSC_vdc}
\end{figure}

\begin{figure}
    \centering
    \includegraphics[width=0.49\columnwidth]{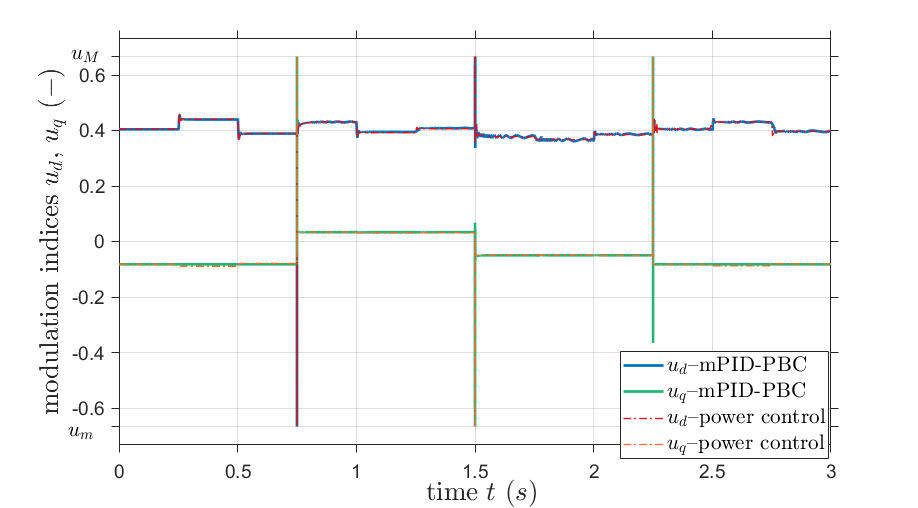}
    \caption{Direct and quadrature components of the modulation indices of the 2L-VSC under mPID-PBC \eqref{eq:PIPBC-I}-\eqref{eq:PIPBC-P}, with $w$ given by \eqref{eq:wboost}, for the considered scenario}
    \label{fig:2LVSC_m}
\end{figure}

For the simulations we consider a point-to-point HVDC transmission system characterized by the parameters given in Table~\ref{table:vsc_parameters}. To validate the mPID-PBC in both nominal and perturbed conditions, we define different practical scenarios over a time span of $12T=3\;s$, and assume that any $T$ an unknown perturbation affects the voltage at the terminal $\mathsf{t2}$. Performances of the controller are further compared with a conventional power PI controller. The voltage nominal value is given by $\hat V_2=775\;kV$ and perturbations are supposed to never exceed $10\%$. Moreover, any $3T$ starting from $0\;s$ a power flow calculation is scheduled. At such time instants, the voltage at the terminal $\mathsf{t2}$ is restored to its nominal value and a new assignable reference vector is provided to the mPID-PBC (nominal conditions). Data about the simulated scenarios are reported in Table~\ref{table:vsc_refs}. Note that these determine power flow reversals at $t={3T,6T}$. For the controller we select the gains as follows: $K_{P}=K_I=\mathbb{I}_2\cdot 10^{-3}$, $K_D=0_{2\times 2}$, $\lambda=0.1$.
\begin{center}
    \begin{table}[t]
    \caption{HVDC transmission system parameters {\cite{HVDCTestGrid}.} }\label{table:vsc_parameters}
    \centering
    \begin{tabularx}{\textwidth}{cc|cc|cc|cc|cc|cc}
\toprule
     $L$ & $\phantom{0}78.2\phantom{0} \;mH$ & $R$ & $0.65\phantom{0} \;\Omega$   &   
     $C$ & $\phantom{0}37.32 \;\mu F$ & $G$ & $0.001 \;mS$ & $V_d$ & $310.27\;kV$ & $\omega$ & $\phantom{00} 50\;Hz$   \\   \hline
      $L_{T1}$ & $120.3 \;mH$ & $R_{T1}$ & $530.96 \;\Omega$ &  $L_{T2}$ & $\phantom{0}60.4  \;mH$ & $R_{T2}$ & $\phantom{0}24.35 \;\Omega$ &
        $L_{T3}$ & $559.6 \;mH$ & $R_{T3}$ & $\phantom{00}3.20 \;\Omega$ \\
    \end{tabularx} 
\end{table}
\end{center}
\begin{table}[t]
\caption{Voltage perturbation ratio at terminal $\mathsf{t2}$ and active and reactive power references in the considered scenario.}\label{table:vsc_refs}
    \centering
    \begin{tabularx}{\textwidth}{c|c|cc|c|cc|c|cc|c|cc}
    \toprule
    & $\mathbf 0$ & $T$ & $2T$& $\mathbf{3T}$& $4T$ & $5T$ & $\mathbf{6T}$ & $7T$& $8T$& $\mathbf{9T}$ & $10T$& $11T$\\
    \hline
     $V_2/\hat{V}_2 (\%)$ & $100 $ & $92 $ & $104$ & $100$ & $110$ & $106$ & $100$ & $103$ & $98$ & $100$ & $94$ & $102$\\
     $P_\star(MW)$  & $+1200$ &   &  & $-480$ &  &   & $+720$ &   &   & $+1200$ &   &  \\
     $Q_\star(MW)$ & $0$ &   &   & $+480$ &   &   & $-360$ &   &   & $0$ &   & 
    \end{tabularx} 
\end{table}
The responses of active and reactive power, the voltages at both terminals and the modulation indices are illustrated in Fig.~\ref{fig:2LVSC_idq}-\ref{fig:2LVSC_m}. It is observed that despite the saturation of the control input at voltage restoration instants, the trajectories of the system quickly converge to a steady-state in all considered scenarios. We also observe that perturbations affecting the terminal $\mathsf{t2}$ determine a steady-state error in terms of active and reactive power, while the voltage at terminal $\mathsf{t1}$ closely follows the variations at terminal $\mathsf{t2}$. Note that it is possible to quantify exactly the normalized regulation error using \eqref{eq:deltaX_vsc}, which in the present case can be shown inferior to $10\%$. These results should be contrasted with the simulated traditional power PI controller, which is instead able to achieve fast and exact regulation independently from the perturbations---at the cost of using an \textit{ad hoc} tuning procedure. Nevertheless, it must be observed that the steady-state errors under PID-PBC are reasonable for several applications (e.g. when some form of storage is available), and the use of such controller come with the benefit of large-signal stability certificates, which strongly simplifies the tuning procedure. Indeed, the PID-PBC does not require the linearization of the system with respect to the wide range of possible operating points.


\section{Conclusions}\label{sec:conclusions}
We have considered the problem of designing PID controllers for general power electronic converters based on passivity arguments, taking in consideration the highly nonlinear dynamics of the underlying circuit. We have established large-signal stability certificates, performance measures and equilibrium-dependent robustness margins for the traditional PID-PBC, showing that these cannot be modified by appropriate tuning of the gains. To overcome this problem, we have thus introduced a leakage in the integral channel that allows to extend robustness margins and performance properties of the controller, at the expense of an approximate regulation---a fact that is captured by the droop characteristic imposed between the control input and the passive output. In addition, we extended our results to the practical scenario where the control input is subject to saturation, by introducing an appropriate monotone modification of the controller. The approach has been validated on two relevant power applications, a lightly damped dc/dc boost converter feeding a constant impedance, constant current (ZI) load and a two-level voltage source converter (2L-VSC) interfaced to an HVDC transmission system.\\
Future research will focus on the straightforward extension to multiterminal dc network, further establishing a precise relation between the choice of the controller parameters and the mode of operation of the power converters in a grid setting. In particular, two research direction seem particularly relevant: first, to evaluate the grid-forming capabilities of the two-level voltage source converter in closed-loop with the PLID-PBC; second, to investigate the application of the proposed  controller to modular multi-level converters (MMCs), which are the current state-of-the art in HVDC transmission systems. Finally, another relevant aspect that will be considered is the definition of a systematic design of the monotone map $w$ and of the controller parameters to improve the closed-loop system performances.
 
 \section*{Author contributions}
D.Z. and G.B. conceived the idea. D.Z. wrote the original manuscript, developed theoretical proofs, performed calculations and simulations for the boost converter application. G.B. provided power electronics-related technical insights and performed calculations and simulations for the HVDC application. R.O. and N.M. contributed to the theoretical proof of Proposition 2. All authors provided critical feedback and helped shape the research, analysis and manuscript. 
 
 \bibliographystyle{unsrt}
\bibliography{biblio}

\end{document}